\documentclass[prb,aps,twocolumn]{revtex4}
\usepackage{amsmath,amssymb,amsfonts,float}
\usepackage[dvips]{graphicx,color}    
\usepackage{times}       
 
\newcommand{\bolr}{\mathbf{r}}

\newcommand{\bra}[1]{\langle #1 |}  
\newcommand{\ket}[1]{| #1 \rangle}  
\newcommand{\VEV}[1]{\langle #1 \rangle}  

\newcommand{\be}{\text{e}}

\newsavebox{\dotdot}
\savebox{\dotdot}[3mm]{\shortstack{\circle*{0.8}\\ \\ \circle*{0.8}}}

\begin{document}
\title{Ground-state phase diagram and magnetic properties of a 
tetramerized spin-$\frac{1}{2}$ $J_1-J_2$ model:\\
BEC of bound
magnons and absence of the transverse magnetization}
\author{H.~T.~Ueda and K.~Totsuka}
\affiliation{Yukawa Institute for Theoretical Physics, 
Kyoto University, Kitashirakawa Oiwake-Cho, Kyoto 606-8502, Japan}
\begin{abstract}
We study the ground state and the magnetization process of 
a spin-1/2 $J_1$-$J_2$ model with a plaquette structure by using 
various methods.  
For small inter-plaquette interaction, this model is expected to 
have a spin-gap and 
we computed the first- and the second excitation energies. 
If the gap of the lowest excitation closes, the corresponding particle 
condenses to form magnetic orders.  
By analyzing the quintet gap and magnetic interactions among 
the quintet excitations, we find a spin-nematic phase around $J_1/J_2\sim -2$ 
due to the strong frustration and the quantum effect.   
When high magnetic moment is applied, not the spin-1 excitations but the spin-2 ones soften and dictate the magnetization process.
We apply a mean-field approximation to the effective Hamiltonian  
to find three different types of phases (a conventional BEC phase, ``striped'' 
supersolid phases and a 1/2-plateau). Unlike the BEC in spin-dimer systems, this BEC phase is not accompanied by transverse magnetization.
Possible connection to the recently discovered spin-gap compound 
$\text{(CuCl)LaNb}_{2}\text{O}_{7}$ is discussed. 
\end{abstract}            
\maketitle
\section{Introduction}
\label{sec:intro}
Magnetic frustration have provided us with many intriguing topics 
e.g. the phenomena of order-by-disorder, the residual entropy at 
absolute zero temperature, disordered spin liquids, etc\cite{Diep-book}. 
There are a variety of models which are known to exhibit the so-called 
frustration effects.
Among them, the $S=1/2$ $J_1$-$J_2$ model on a square lattice 
has been extensively investigated over the last two decades 
as one of the simplest models to study how frustration destroys 
magnetic orders and stabilizes paramagnetic phases.  
The model is defined by adding antiferromagnetic interactions 
on diagonal bonds to the ordinary Heisenberg antiferromagnet 
on a square lattice (see FIG.~\ref{fig:plafull}): 
\begin{equation}
{\cal H} = J_{1}\sum_{\text{n.n.}}{\bf S}_{i}{\cdot}{\bf S}_{j}
+ J_{2}\sum_{\text{n.n.n.}}{\bf S}_{i}{\cdot}{\bf S}_{j} \; ,
\end{equation}
where the summations (n.n.) and (n.n.n.) are taken for 
the nearest-neighbor- and the second-neighbor (diagonal) pairs, 
respectively.   
In the classical ($S\nearrow \infty$) limit, the ground state 
is readily obtained by computing Fourier transform $J(\boldsymbol{k})$ 
of the exchange interactions and minimize it in the 
$\boldsymbol{k}$-space: 
\begin{itemize}
\item $J_{1}>0$, $J_{2}< J_{1}/2$: the ground state has N\'{e}el 
antiferromagnetic order (NAF).
\item $J_{2}>|J_{1}|/2$: the ground state consists of two 
interpenetrating N\'{e}el-ordered square lattices. 
First quantum correction fixes the relative angle between 
the two ordering directions and selects the so-called {\em collinear 
antiferromagnetic order} (CAF). 
\item otherwise: the ferromagnetic (FM) ground state is stabilized. 
\end{itemize}
For $J_{2}<0$, the next-nearest-neighbor (diagonal) interaction 
gives rise to no frustration and only the case with $J_{2}>0$ 
is non-trivial.  The case $J_{1}, J_{2}>0$ has been extensively 
studied in the context of spin-gap phases stabilized by 
the frustrating interactions.  
Chandra and Doucot\cite{Chandra-D-88} investigated the model 
in the large-$S$ limit and concluded that a non-magnetic 
(neither NAF nor CAF) phase appeared around the classical 
phase boundary $J_{2}/J_{1}=1/2$. The most quantum case $S=1/2$ has been 
studied later both by numerical\cite{Schulz-Z-96,Capriotti-01} and by 
analytical methods \cite{Singh-W-H-O-99,Kotov-00} 
(for other literatures, see, for instance, 
Refs. \onlinecite{Misguich-L-03,Oitmaa-book06} and references cited therein). 
By now it is fairly well established that we have spin 
gapped phase(s) in the window $0.4\lesssim J_{2}/J_{1} \lesssim 0.6$ 
although the nature of the spin-gap phase(s) is still in controversy.   

The case with $J_{1}<0$, $J_{2}>0$ has been less investigated 
and recent analyses\cite{Shannon-04,Shannon-06} suggested that 
there is another non-magnetic (probably spin-nematic) phase 
around the classical boundary $J_{2}/J_{1}=-1/2$ between CAF and FM.  
From an experimental viewpoint, most compounds\cite{Melzi-00,Kaul-03} 
found so far correspond to the ordered phase (CAF) of the $J_{1}$-$J_{2}$ 
model.  

Recently, Kageyama {\em et al.} reported\cite{Kageyama-05-1} 
a new two-dimensional Cu-based compound (CuCl)La$\text{Nb}_{2}\text{O}_{7}$.  
In this compound, two-dimensional sheets 
consisting of $\text{Cu}^{2+}$ and $\text{Cl}^{-}$ are separated 
from each other by non-magnetic [La$\text{Nb}_{2}\text{O}_{7}$] layers 
and within each sheet the $\text{Cu}^{2+}$ ions form 
a square lattice.  The $\text{Cl}^{-}$ ions are located 
at the center of plaquettes and from a naive Goodenough-Kanamori 
argument the $S=1/2$ $J_{1}$-$J_{2}$ model with 
$J_{1}<0$ and $J_{2}>0$ is suggested as the model Hamiltonian 
for (CuCl)La$\text{Nb}_{2}\text{O}_{7}$.  

What is remarkable with this compound is that 
inelastic neutron scattering experiments \cite{Kageyama-05-1} 
observed a finite spin gap 2.3meV(=26.7K) above the spin-singlet 
ground state. 
Subsequently, high-field magnetization measurements\cite{Kageyama-05-2} 
were carried out to show that magnetization monotonically increased 
between two critical fields $H_{\text{c1}}=10.3$T and 
$H_{\text{c2}}=30.1$T.   The data for (i) the Weiss temperature 
and (ii) the saturation field $H_{\text{c2}}$ in principle determine 
the coupling constants $J_{1}$ and $J_{2}$.  
Unfortunately, none of the solutions $(J_{1},J_{2})$ 
obtained in this way reproduced the spin-gap behavior\cite{Kageyama-05-2}.  
Therefore, the usual $J_{1}$-$J_{2}$ model does not seem to work.  

The second intriguing point concerns the magnetization process.  
From the standard scenario \cite{Nikuni-O-O-T-00}, 
the onset of magnetization at $H=H_{\text{c1}}$ in spin-gapped 
systems is understood as Bose-Einstein condensation 
(BEC, or superfluid onset, more precisely) of the lowest-lying 
triplet excitation (magnon) and the lower critical 
field $H_{\text{c1}}$ at $T=0$ is given by the spin gap $\Delta$ as 
$H_{\text{c1}}=\Delta/(g\mu_{\text{B}})$.  
This BEC scenario has been confirmed in various spin gap 
compounds\cite{Tanaka-01,Han-purple,Stone-06}.  
 
Recent specific-heat- and magnetization measurements \cite{Kitada-07} 
for (CuCl)La$\text{Nb}_{2}\text{O}_{7}$ 
exhibited behavior typical of spin-BEC transitions and suggested that 
the magnetization-onset transition at $H_{\text{c1}}$ may be described 
by BEC of a certain kind of magnetic excitations.  
However, we immediately find a serious difficulty when we try to 
understand this within the standard BEC scenario; 
the lower critical field $H_{\text{c1}}=18.4$T expected 
from the observed spin gap $\Delta=2.3$meV at the zero field
(where the experimental value $g=2.17$ is used) 
in the standard scenario is much larger than the observed 
value\cite{Kageyama-05-2} $H_{\text{c1}}=10.3$T. 
One possible explanation for this discrepancy may be that 
a lower-lying triplet excitation which is responsible for the BEC was not 
observed in the neutron-scattering experiments because of 
selection rules.   However, this seems unlikely since powder samples 
were used and usually one can hardly expect a perfect extinction 
of a certain triplet excitation in such powder samples.  
Neither susceptibility measurements \cite{Kageyama-05-1} 
nor NMR data \cite{Yoshida-07} indicate such a hidden triplet  
excitation.    

An alternative and a more appealing scenario would be that 
the BEC occurs not in a single-particle channel but in a multi-particle 
channel.  That is, what condenses to support a spin-superfluid 
is a bound state of magnon excitations.   
The possibility of multi-magnon condensation has been proposed 
theoretically \cite{Momoi-Totsuka,Totsuka-M-U-01} 
in the context of a {\em kinetic} quintet bound state in 
the Shastry-Sutherland model (see Ref.\onlinecite{Miyahara-U-review} 
and references cited therein).  
In fact, gapped quintet excitations which come down as the external field 
is increased were observed in the ESR experiments \cite{Nojiri-03} 
carried out for Sr$\text{Cu}_{2}(\text{BO}_{3})_{2}$, whereas 
small Dzyaloshinskii-Moriya interactions hindered a quintet BEC from 
being observed in that compound (see also Ref.\onlinecite{RBF}).    

One of the simplest $J_{1}$-$J_{2}$-like models which realize 
the above scenario and have a finite spin gap 
would be the $S=1/2$ $J_{1}$-$J_{2}$ model 
with a plaquette structure (see FIG.~\ref{fig:plafull}).  
A similar model ($J_{1},J_{2}>0$) has been investigated to develop 
a plaquette series expansion\cite{Singh-W-H-O-99}.  
In this paper, we mainly focus on the region 
$J_{1}<0$, $J_{2}>0$ where the quintet excitation is expected 
to play an important role in low-energy physics.  

The organization of the present paper is as follows.  
In section II, we briefly recapitulate the problem of a single plaquette
mainly to establish the  notations. 
The coupling among plaquettes will be taken into account in 
section \ref{sec3} by two different methods: 
(i) a plaquette extension of the bond-operator mean-field 
theory\cite{Sach-Bhatt} and 
(ii) a perturbation expansion with respect to the inter-plaquette 
couplings.   We find gapped triplets and quintet for small enough 
inter-plaquette couplings in both methods.  

For larger values of inter-plaquette couplings, one of the gapped 
excitations softens and the form of the effective interactions among 
the soft excitations determines the resulting magnetic phases.  
By using the gaps obtained in the perturbation expansion, 
we determine the semi-quantitative phase diagram in section 
\ref{sec:phase} (see FIG.~\ref{fig:souzuG} and FIG.~\ref{fig:nf}).   

The effect of high magnetic field will be considered 
in section \ref{sec:magnet}.  
For high enough field compared with the spin gaps, we can approximate 
the low-energy sector by using only the singlet and the lowest 
excited state.  For $J_{1}<0$, we may expect that the quintet touches 
the singlet ground state first and a multi-particle BEC occurs. On general grounds, a single-particle (magnon) BEC phase is expected to have finite transverse magnetization. Actually, in the BEC phase of TlCuCl$_3$, the transverse magnetization has been observed in the experiment\cite{Tanaka-01}. In the case of a multi-particle BEC, however, the transverse magnetization does not appear.
To investigate the magnetization process, we shall keep only the singlet 
and the quintet to derive a hardcore boson model 
as the effective Hamiltonian valid in high enough magnetic field.  
A mean-field approximation\cite{TY} will be applied 
to the resulting effective Hamiltonian to draw a full magnetization curve.  
Interesting phases (a 1/2-plateau and supersolids) will be discussed. 
According to the value of the parameters, we shall roughly classify 
the magnetization curve in FIG.~\ref{Fig:magsouzu}.  

A summary of the main results and the discussion on the connection 
to the spin-gap compound $\text{(CuCl)LaNb}_{2}\text{O}_{7}$ 
will be given in sections \ref{sec:CuCl} and \ref{sec:Conclusion}, 
respectively.   
The equations omitted in the text will be summarized in the appendices.      
\section{plaquette structure}
\label{sec:section-2}
\begin{figure}[ht]
\begin{center}
\includegraphics[scale=0.45]{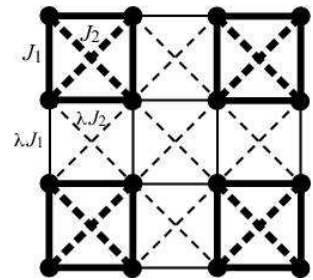}
\caption{Two dimensional square lattice with a plaquette structure 
to be considered in this paper. 
Filled circles denote spin-1/2s connected by the usual exchange
interactions. 
Thin lines (both solid and broken) imply that the interactions are 
multiplied by the distortion parameter $\lambda$ on 
these bonds.\label{fig:plafull}}
\end{center}
\end{figure}

\begin{figure}[ht]
\begin{center}
\includegraphics[scale=0.45]{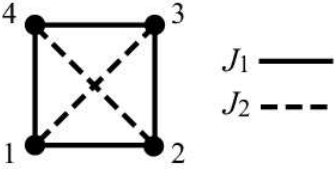}
\caption{Single plaquette. Dots represent spin 1/2s, 
and the solid- and the dashed lines respectively represent 
Heisenberg interaction with the couplings $J_1$ and $J_2$ 
between $S=1/2$ spins.\label{fig:plas}}
\end{center}
\end{figure}

We consider a spin-1/2 $J_1$-$J_2$ model with a plaquette structure 
where the interactions among spin-1/2s are explicitly tetramerized 
(see FIG.~\ref{fig:plafull}).  
The model is made up of four-spin units ({\em plaquettes}) 
and the four sites constituting a single plaquette are 
connected by the nearest-neighbor- ($J_{1}$) and 
the second-neighbor ($J_{2}$) interactions as is shown in 
FIG.~\ref{fig:plas}). 
The inter-plaquette interactions (both the nearest-neighbor- and 
the diagonal) which connect those units are multiplied 
by a distortion constant $\lambda$ $(0 \leq \lambda \leq 1)$.  
This parameter may be thought of as modeling 
the distortion of the underlying lattice in a simple way.  
In the case of $\lambda =1$, 
this model reduces to the homogeneous $J_1$-$J_2$ model, 
while when $\lambda =0$, the plaquettes are decoupled from each other.  
\subsection{single plaquette}
\begin{figure}[ht]
\begin{center}
\includegraphics[scale=1]{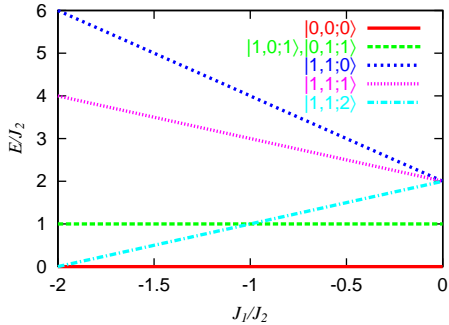}
\caption{The energy of the triplets $\ket{1,0;1}$, $\ket{0,1;1}$ 
and the quintet $\ket{1,1;2}$. 
We take the units of energy as $J_2$, and the energy is plotted 
as a function of $J_1/J_2$\label{fig:Elebel}}
\end{center}
\end{figure}

Let us begin by analyzing a single isolated plaquette, which 
corresponds to the case $\lambda=0$.  
The eigenstates of a single plaquette can be easily obtained as
follows. 
First we note that a plaquette Hamiltonian can be rewritten as
\begin{equation}
\begin{split}
H =&  J_{1}({\bf S}_{1}{\cdot}{\bf S}_{2}
+{\bf S}_{2}{\cdot}{\bf S}_{3} + {\bf S}_{3}{\cdot}{\bf S}_{4}
+ {\bf S}_{4}{\cdot}{\bf S}_{1}) \\
& + J_{2}( {\bf S}_{1}{\cdot}{\bf S}_{3}
+{\bf S}_{2}{\cdot}{\bf S}_{4}) \\
 = & \frac{J_{1}}{2}{\bf S}^{2} + \frac{1}{2}(J_{2}-J_{1})
({\bf S}_{a}^{2}+{\bf S}_{b}^{2})
-\frac{3}{2} J_{2}\ ,
\label{Hpsingle}\
\end{split}
\end{equation}
where ${\bf S}_{a}={\bf S}_{1}+{\bf S}_{3}$,
${\bf S}_{b}={\bf S}_{2}+{\bf S}_{4}$ and 
${\bf S}={\bf S}_{a}+{\bf S}_{b}$.
Therefore, all the $2^{4}$ eigenstates are classified by 
the three quantum numbers as $|S_{a},S_{b};S \rangle$. 
The eigenvalues $E(S_{a},S_{b},S)$ are given by
\begin{subequations}
\begin{align}
& E(0,0,0)=0\label{E0}\ ,\\
& E(1,1,0)=2J_{2}-2J_{1}\ ,\\
& E(1,0,1)=E(0,1,1)=J_{2}\ ,\label{Et}\\
& E(1,1,1)=-J_{1}+2J_{2}\ ,\\
& E(1,1,2)=J_{1}+2J_{2}\ .\label{Eq}
\end{align}
\label{sec2:EE}
\end{subequations}
Here a constant $-\frac{3}{2} J_{2}$ has been dropped just for simplicity. 
The energy of these states is shown in FIG.~\ref{fig:Elebel}.  
For $-1 < J_1 /J_2 < 0$, the spin-singlet state $\ket{0,0;0}$ 
is the ground state, the triplets $\ket{1,0;1}$, $\ket{0,1;1}$ are 
the first excited states, and the quintet $\ket{1,1;2}$ is 
the second excited state.  
For $-2 < J_1 /J_2 < -1$, the singlet $\ket{0,0;0}$ is the ground state, 
quintet $\ket{1,1;2}$ is the first excited state, 
and triplets $\ket{1,0;1}$, $\ket{0,1;1}$ are the second excited state. \\
The singlet $\ket{0,0;0}$ is written as
\begin{equation}
\begin{split}
|s\rangle & \equiv |0,0;0\rangle\\
& =\frac{1}{2}(|\uparrow \downarrow\rangle -|\downarrow \uparrow\rangle)
(|\uparrow \downarrow\rangle -|\downarrow \uparrow\rangle )  \; .
\label{ss1}
\end{split}
\end{equation}
In what follows, the single-spin states in ket will be shown 
in the order of 1,3,2,4, i.e. $|s_{1},s_{3},s_{2},s_{4}\rangle$ 
in FIG.~\ref{fig:plas}.  
For later convenience, we name the two triplets $\ket{1,0;1}$ 
and $\ket{0,1;1}$ as $\ket{p_i}$ and $\ket{q_i}$ ($i=x,y,z$)
respectively.  
The explicit expressions of the two triplets are given as:
\begin{subequations}
\begin{align}
|p_{x}\rangle & =-\frac{1}{2}(|\uparrow \uparrow\rangle 
-|\downarrow \downarrow\rangle )(|\uparrow \downarrow\rangle 
-|\downarrow \uparrow\rangle )\ ,\\
|p_{y}\rangle & =\frac{i}{2}(|\uparrow \uparrow\rangle 
+|\downarrow \downarrow\rangle )(|\uparrow \downarrow\rangle 
-|\downarrow \uparrow\rangle )\ ,\\
|p_{z}\rangle & =\frac{1}{2}(|\uparrow \downarrow\rangle 
+|\downarrow \uparrow\rangle )(|\uparrow \downarrow\rangle 
-|\downarrow \uparrow\rangle )\ ,
\end{align}
\end{subequations}
\begin{subequations}
\begin{align}
|q_{x}\rangle & =-\frac{1}{2}(|\uparrow \downarrow\rangle  
-|\downarrow \uparrow\rangle )(|\uparrow \uparrow\rangle 
-|\downarrow \downarrow\rangle )\ ,\\
|q_{y}\rangle & =\frac{i}{2}(|\uparrow \downarrow\rangle 
-|\downarrow \uparrow\rangle )(|\uparrow \uparrow\rangle 
+|\downarrow \downarrow\rangle )\ ,\\
|q_{z}\rangle & =\frac{1}{2}(|\uparrow \downarrow\rangle 
-|\downarrow \uparrow\rangle )(|\uparrow \downarrow\rangle 
+|\downarrow \uparrow\rangle )\ .
\end{align}
\end{subequations}

To label the quintet $\ket{1,1;2}$ states, we use the eigenvalues 
of $S^z$, i.e. $\ket{1,1;S{=}2,S^z}$ whose expressions  are given 
explicitly as: 
\begin{subequations}
\begin{align}
&|1,1;2,2\rangle =|\uparrow \uparrow \uparrow \uparrow \rangle\ ,\\
&|1,1;2,1\rangle = \frac{1}{2}\left\{ |\uparrow \uparrow \rangle 
(|\uparrow \downarrow\rangle +|\downarrow \uparrow \rangle )
+(|\uparrow \downarrow \rangle +|\downarrow \uparrow \rangle)
|\uparrow \uparrow \rangle \right\}\ ,\\
&|1,1;2,0\rangle  = \hspace{-0.8mm}\frac{1}{\sqrt{6}}
\left\{ (|\uparrow \downarrow \rangle \hspace{-0.5mm}
+\hspace{-0.5mm}|\downarrow \uparrow \rangle)(|\uparrow \downarrow
\rangle \hspace{-0.5mm}
+\hspace{-0.5mm}|\downarrow \uparrow \rangle)\hspace{-0.5mm}
+\hspace{-0.5mm}
|\hspace{-1mm}\uparrow \uparrow \downarrow \downarrow \rangle
\hspace{-0.5mm}
+\hspace{-0.5mm}|\hspace{-0.8mm}\downarrow \downarrow \uparrow \uparrow 
\rangle \hspace{-0.5mm} \right\},\\
&|1,1;2,-1\rangle  = \frac{1}{2} \{ |\downarrow \downarrow\rangle 
(|\uparrow \downarrow \rangle +|\downarrow \uparrow \rangle)
+(|\uparrow \downarrow \rangle +|\downarrow \uparrow \rangle)
|\downarrow \downarrow \rangle \}\ ,\\
&|1,1;2,-2\rangle  = |\downarrow \downarrow \downarrow \downarrow \rangle \ .
\end{align}
\label{qq}
\end{subequations}
\section{Effect of inter-plaquette interaction}
\label{sec3}
For $\lambda=0$ and $J_1/J_2 >-2$, all plaquettes are 
in the singlet state $\ket{0,0;0}$.   
Finite inter-plaquette interactions $\lambda$ induce various tunneling 
processes among plaquettes to change both the ground state 
and the excitations over it.  
For finite $\lambda$, we calculate the excitation energy 
by two different approaches.  
One is the bond-operator mean-field theory (MFT)\cite{Sach-Bhatt,Zhito-Ueda}, 
which gives the excitation energy of the triplets $\ket{p_i}$,
$\ket{q_i}$. 
Another is the second-order perturbation theory in $\lambda$, 
and it gives the energy of the quintet $\ket{1,1;2}$ as well as 
that of $\ket{p_i}$ and $\ket{q_i}$ . For sufficiently small $\lambda$, 
both approximations yield finite energy gaps for these excitations 
and when one of these gaps closes, the corresponding (bosonic) 
excitation condenses to form a magnetically ordered state.  
The energy of triplet excitations can be observed by 
inelastic neutron scattering experiments.  
Both approximations may not be reliable for large $\lambda$ 
and small $|J_1/J_2|$.  
\subsection{bond-operator MFT}
Let us begin with the bond-operator MFT\cite{Sach-Bhatt,Zhito-Ueda}. 
For $-2 < J_1 /J_2 < 0$ and $\lambda=0$, $\ket{0,0;0}$ is the ground state 
and the degenerate triplets $\ket{1,0;1}$, $\ket{0,1;1}$ are 
the first- or the second excited state (see FIG.~\ref{fig:Elebel}).  
Therefore, we may truncate the Hilbert space and consider a subspace 
spanned by the singlet $\ket{0,0;0}$ and the triplets 
$\ket{p_i}$, $\ket{q_i}$.  
This approximation is reliable to estimate the excitation energy 
of the triplets, unless $\ket{1,1;2}$ condenses.  
In this subspace, nonzero matrix elements of ${\bf S}_{1,2,3,4}$ is 
\begin{subequations}
\begin{align}
& \langle s|S_{1}^{\alpha}|p_{\beta}\rangle =\frac{1}{2}
\delta_{\alpha\beta}\label{s1}\ ,\ & \langle p_{\alpha}
|S_{1}^{\beta}|p_{\gamma}\rangle =\frac{i}{2}
\epsilon_{\alpha \beta \gamma}\ ,\\
& \langle s|S_{2}^{\alpha}|q_{\beta}\rangle 
=\frac{1}{2}\delta_{\alpha\beta}\label{s2}\ ,\ & 
\langle q_{\alpha}|S_{2}^{\beta}|q_{\gamma}\rangle 
=\frac{i}{2}\epsilon_{\alpha \beta \gamma}\ ,\\
& \langle s|S_{3}^{\alpha}|p_{\beta}\rangle 
=-\frac{1}{2}\delta_{\alpha\beta}\label{s3}\ ,\ & 
\langle p_{\alpha}|S_{3}^{\beta}|p_{\gamma}\rangle 
=\frac{i}{2}\epsilon_{\alpha \beta \gamma}\ ,\\
& \langle s|S_{4}^{\alpha}|q_{\beta}\rangle 
=-\frac{1}{2}\delta_{\alpha\beta}\label{s4}\ ,\ & 
\langle q_{\alpha}|S_{4}^{\beta}|q_{\gamma}\rangle 
=\frac{i}{2}\epsilon_{\alpha \beta \gamma}\ ,
\end{align}
\label{psgS}
\end{subequations}
where $\alpha,\beta\hspace{-2mm}=\hspace{-2mm}x,y,z$.  
Using boson operators $s, p_\alpha, q_\alpha(\alpha=x,y,z)$ 
satisfying the standard commutation relations, 
$[s,s^\dagger ]=1$, $[p_\alpha ,p_\beta ^\dagger]=\delta_{\alpha\beta}$,
$[q_\alpha ,q_\beta ^\dagger]=\delta_{\alpha\beta}$, 
$[s, p_\alpha]=0$, etc, the local spin operator 
${\bf S}_{1,2,3,4}$ can be written as
\begin{subequations}
\begin{align}
& S_{1}^{\alpha}=\frac{1}{2}(s^{\dagger}p_{\alpha}+s
 p_{\alpha}^{\dagger})
-\frac{i}{2}\epsilon_{\alpha\beta\gamma}p^\dagger_\beta p_\gamma\ ,
\label{bt1}\\
&S_{3}^{\alpha}=-\frac{1}{2}(s^{\dagger}p_{\alpha}
+s p_{\alpha}^{\dagger})-\frac{i}{2}
\epsilon_{\alpha\beta\gamma}p^\dagger _\beta p_\gamma\ ,\\
&S_{2}^{\alpha}=\frac{1}{2}(s^{\dagger}q_{\alpha}
+s q_{\alpha}^{\dagger})-\frac{i}{2}
\epsilon_{\alpha\beta\gamma}q^\dagger _\beta q_\gamma\ ,\\
&S_{4}^{\alpha}=-\frac{1}{2}(s^{\dagger}q_{\alpha}
+s q_{\alpha}^{\dagger})-\frac{i}{2}
\epsilon_{\alpha\beta\gamma}q^\dagger _\beta q_\gamma\ ,\label{bt4}
\end{align}
\end{subequations}
where the summation over repeated indices is implied.  
Since the restriction that each plaquette has exactly one particle leads 
to the local constraint 
$s^\dagger s+\sum_\alpha (p^\dagger_\alpha p_\alpha 
+q^\dagger_\alpha q_\alpha)=1$, we introduce the Lagrange multiplier 
$\mu_i$ and add a term 
\begin{equation}
\mu_{i}\left\{s_{i}^\dagger s_{i}
+\sum_\alpha (p^\dagger_{\alpha,i} p_{\alpha,i}
+q^\dagger_{\alpha,i} q_{\alpha,i})-1 \right\} \; .
\end{equation} 
to each plaquette Hamiltonian. 
We may assume that $\mu_i$ for each plaquette 
takes the same value $\mu$ for all plaquettes 
because of the translation invariance. 

Next, we replace $s$ by its expectation value 
$\VEV{s} =\overline{s}$, since the $s$ boson  
condenses in the ground state.  Moreover, since the triplet is dilute 
when the energy gap is positive, we may ignore the terms consisting 
of three or four triplet operators.  
In this way, we obtain the mean-field Hamiltonian $H_{\text{bo}}$ 
consists only of bilinear terms in $p$ and $q$.  
The mean-field parameters $(\mu,\overline{s})$ are determined by
requiring the expectation values of the derivatives of $H_{\text{bo}}$ 
with respect to the mean-field (MF) ground state vanish:
\begin{equation}
\Bigl\langle \frac{\partial H_{\text{bo}}}{\partial \mu}
\Bigr\rangle _{\text{MF}} =0
\; , \quad 
\Bigl\langle \frac{\partial H_{\text{bo}}}{\partial \overline{s}}
\Bigr\rangle_{\text{MF}} =0\ ,
\end{equation}
or equivalently by finding the extrema of the mean-field ground-state 
energy $E_{\text{G.S.}}^{\text{mf}}$:
\begin{equation}
\frac{\partial E_{\text{G.S.}}^{\text{mf}}}{\partial \mu}=0 \; , \quad 
\frac{\partial E_{\text{G.S.}}^{\text{mf}}}{\partial \overline{s}}=0 \; .
\label{eqn:variational}
\end{equation}
In particular, $E_{\text{G.S.}}^{\text{mf}}$ must be minimum 
for $\overline{s}$. 

\begin{figure}[ht]
\begin{center}
\includegraphics[scale=0.6]{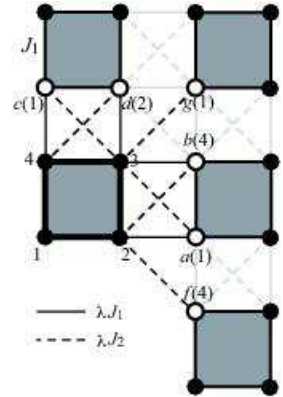}
\caption{Inter-plaquette interactions associated with the plaquette 
$n$ (shown by a thick line).\label{fig:heizu}}
\end{center}
\end{figure}

In this approximation, the inter-plaquette interactions 
associated with the site $n$ reads
\begin{subequations}
\begin{align}
& (H_{\hat{x}})_n 
=J_1 ({\bf S}_2 {\cdot} {\bf S}_a +{\bf S}_3 {\cdot} {\bf S}_b)
+J_2({\bf S}_3 {\cdot} {\bf S}_a +{\bf S}_2{\cdot} {\bf S}_b) \notag \\
& = \frac{J_1 }{4}\overline{s}^2
\{(q_\alpha \hspace{-0.8mm}+\hspace{-0.5mm}q^\dagger_\alpha)_n 
(p_\alpha\hspace{-0.8mm}+\hspace{-0.5mm}p_\alpha ^\dagger)_{n+\hat{x}}
\hspace{-0.5mm}+\hspace{-0.5mm} 
(p_\alpha \hspace{-0.8mm}+\hspace{-0.5mm}p_\alpha ^\dagger)_{n}
(q_\alpha \hspace{-0.8mm}+\hspace{-0.5mm}q^\dagger_\alpha)_{n+\hat{x}}\} 
\nonumber\\
& -\frac{J_2 }{4}\overline{s} ^2\{(p_\alpha \hspace{-0.8mm}
+\hspace{-0.5mm}p_\alpha ^\dagger)_{n}(p_\alpha \hspace{-0.8mm}
+\hspace{-0.5mm} p_\alpha ^\dagger)_{n+\hat{x}} \hspace{-0.5mm}
+\hspace{-0.5mm}(q_\alpha \hspace{-0.8mm}
+\hspace{-0.5mm}q^\dagger_\alpha)_n (q_\alpha \hspace{-0.8mm}
+\hspace{-0.5mm} q^\dagger_\alpha)_{n+\hat{x}}\} ,\\
& (H_{\hat{y}})_n 
=J_1 ({\bf S}_3 {\cdot} {\bf S}_d +{\bf S}_4{\cdot} {\bf S}_c)
+J_2({\bf S}_3{\cdot} {\bf S}_c +{\bf S}_4{\cdot} {\bf S}_d) \notag \\
& =\hspace{-0.7mm}-\frac{J_1 }{4}\overline{s}^2
\{ \hspace{-0.5mm}(q_\alpha \hspace{-0.8mm}+\hspace{-0.5mm}
q^\dagger_\alpha)_n (p_\alpha \hspace{-0.8mm}+\hspace{-0.5mm}
p_\alpha^\dagger)_{n+\hat{y}}\hspace{-0.8mm}+\hspace{-0.5mm} 
(p_\alpha \hspace{-0.8mm}+\hspace{-0.5mm}p_\alpha ^\dagger)_{n}
(q_\alpha \hspace{-0.8mm}+\hspace{-0.5mm}q^\dagger_\alpha)_{n+\hat{y}} 
\hspace{-0.5mm}\} 
\nonumber\\
& -\frac{J_2 }{4}\overline{s}^2\{(p_\alpha \hspace{-0.8mm}
+\hspace{-0.5mm}p_\alpha ^\dagger)_{n}(p_\alpha \hspace{-0.8mm}
+\hspace{-0.5mm}p_\alpha ^\dagger)_{n+\hat{y}} \hspace{-0.5mm}
+\hspace{-0.5mm}(q_\alpha \hspace{-0.8mm}+\hspace{-0.5mm}
q^\dagger_\alpha)_n (q_\alpha \hspace{-0.8mm}
+\hspace{-0.5mm}q^\dagger_\alpha)_{n+\hat{y}} \} ,\\
& (H_{\hat{x}+\hat{y}})_n=J_2 {\bf S}_3{\cdot} {\bf S}_g \nonumber\\
& \ \ =-\frac{J_2 }{4}\overline{s} ^2 (p_\alpha +p_\alpha^\dagger)_{n}
(p_\alpha +    p_\alpha ^\dagger)_{n+\hat{x}+\hat{y}} , \\
& (H_{\hat{x}-\hat{y}})_n=J_2 {\bf S}_2{\cdot} {\bf S}_f \nonumber\\
& \ \ =-\frac{J_2 }{4}\overline{s} ^2 (q_\alpha +q^\dagger_\alpha)_n 
(q_\alpha +q^\dagger_\alpha)_{n+\hat{x}-\hat{y}}\ ,
\end{align}
\end{subequations}
where the site labels $1,\dots,4$ and $a,\dots,g$ are defined 
in FIG.~\ref{fig:heizu}. 

Summing up all four interactions and doing Fourier transformation, 
the total Hamiltonian $H_{\text{bo}}$ reads
\begin{subequations}
\begin{align}
H_{\text{bo}}
=&\sum_{{\bf k}}\Bigl\{ J_2 (p^{\dagger \alpha}_{{\bf k}}p_{{\bf k}}^\alpha
+ q^{\dagger \alpha}_{{\bf k}}q_{{\bf k}}^\alpha)\nonumber\\
& -\frac{J_2}{4}\overline{s}^2 f_{+}({\bf k})
(p^{\dagger \alpha}_{{\bf k}}p_{{\bf k}}^\alpha 
+p^{\alpha}_{{\bf k}}p_{{\bf k}}^{\dagger\alpha}+p^{\alpha}_{{\bf k}}
p_{-{\bf k}}^\alpha+p^{\dagger \alpha}_{{\bf k}}
p_{-{\bf k}}^{\dagger\alpha})\notag \\
& -\frac{J_2}{4}\overline{s}^2 f_{-} ({\bf k})
(q^{\dagger \alpha}_{{\bf k}}q_{{\bf k}}^\alpha
+q^{\alpha}_{{\bf k}}q_{{\bf k}}^{\dagger\alpha}
+q^{\alpha}_{{\bf k}}q_{-{\bf k}}^\alpha
+q^{\dagger \alpha}_{{\bf k}}q_{-{\bf k}}^{\dagger\alpha})\notag \\
& +\frac{J_1}{2}\overline{s}^2(\cos k_x-\cos k_y)
(p^{\dagger\alpha}_{{\bf k}}q^{\alpha}_{{\bf k}}
+p^{\alpha}_{{\bf k}}q^{\alpha}_{-{\bf k}}+h.c.)\notag \\
& -\mu(\overline{s} ^2+p^{\dagger\alpha}_{{\bf k}}
p^{\alpha}_{{\bf k}}+q_{{\bf k}}^{\dagger\alpha}q_{{\bf k}}^\alpha-1)
\Bigr\}  \; ,
\end{align}
\end{subequations}
where we have defined 
\begin{equation}
f_{\pm}({\bf k})\equiv\cos k_{x}+\cos k_{y}+\cos(k_{x}\pm k_{y})\ .
\end{equation}
If we introduce a vector 
${\bf v}_{{\bf k}} =(
p_{{\bf k}}\ ,\ q_{{\bf k}}\ ,\ p^{\dagger}_{-{\bf k}}\ ,\ 
q^{\dagger}_{-{\bf k}}) ^{\text{T}}$, 
the MF Hamiltonian $H_{\text{bo}}$ can be written compactly as
\begin{equation}
H_{\text{bo}}=\sum_{{\bf k}}
{\bf v}^{\dagger\alpha}_{{\bf k}} A({\bf k}){\bf v}^\alpha_{{\bf k}} 
-N_{\text{p}}\left\{ 3(J_2-\mu)+\mu (\overline{s} ^2 -1)\right\}\ ,\label{vav}
\end{equation}
where $N_{\text{p}}$ denotes the total number of plaquettes and the kernel 
$A({\bf k})$ is given as
\begin{subequations}
\begin{align}
A({\bf k}) &=\left(
\begin{array}{cccc}
a_{\text{bo}} ({\bf k}) & b_{\text{bo}} ({\bf k}) & c_{\text{bo}}({\bf k}) 
& d_{\text{bo}} ({\bf k})\\
b_{\text{bo}} ({\bf k}) & e_{\text{bo}} ({\bf k}) & b_{\text{bo}} ({\bf k}) 
& d_{\text{bo}} ({\bf k})\\
c_{\text{bo}} ({\bf k}) & b_{\text{bo}} ({\bf k}) & a_{\text{bo}} ({\bf k}) 
& b_{\text{bo}} ({\bf k})\\
b_{\text{bo}} ({\bf k}) & d_{\text{bo}} ({\bf k}) & b_{\text{bo}} ({\bf k}) 
& e_{\text{bo}} ({\bf k})\\
\end{array}
\right) \ ,
\end{align}
\end{subequations}
\begin{subequations}
\begin{align}
a_{\text{bo}} ({\bf k}) 
&=\frac{J_2}{2}-\frac{J_2}{4}\overline{s}^2 f_{+} ({\bf k})-\frac{\mu}{2}\ ,\\
b_{\text{bo}} ({\bf k}) 
&= -\frac{J_1}{4}\overline{s}^2 (\cos k_x -\cos k_y)\ ,\\
c_{\text{bo}} ({\bf k}) &= -\frac{J_2}{4}f_+ ({\bf k})\ ,\\
d_{\text{bo}} ({\bf k}) &= -\frac{J_2}{4}f_- ({\bf k})\ ,\\
e_{\text{bo}} ({\bf k}) 
&= \frac{J_2}{2}-\frac{J_2}{4}\overline{s}^2 f_- ({\bf k})-\frac{\mu}{2}\ .
\end{align}
\end{subequations}
Using a $4\times 4$ real matrix $L_{\bf k}$ 
(see Appendix~\ref{sec:appendix-1} for the detail), 
we can diagonalize $A({\bf k})$ by the Bogoliubov transformation:
\begin{equation}
\begin{split}
& L_{{\bf k}}{\bf v}_{{\bf k}}={\bf v}^{\prime}_{{\bf k}}\; , \\
& {\bf v}_{{\bf k}}^{\prime} =\left(
\begin{array}{c}
p_{{\bf k}}^\prime\ ,\ q_{{\bf k}}^{\prime}\ ,\ p^{\dagger\prime}_{-{\bf k}}\ ,\ q^{\dagger\prime}_{-{\bf k}}\\
\end{array}
\right) ^{\text{T}}\ .\label{LL}
\end{split}
\end{equation}
As is shown in Appendix~\ref{sec:appendix-1}, $H_{\text{bo}}$ then reduces to
\begin{equation}
H_{{\rm bo}}=\sum_{{\bf k}}\left\{
\omega_1({\bf k})p^{\prime\dagger\alpha}_{{\bf k}}p^{\prime\alpha}_{{\bf k}}
+\omega_2({\bf k}) q^{\prime\dagger\alpha}_{{\bf k}}q^{\prime\alpha}_{{\bf k}}\right\}
+E^{\text{mf}}_{\text{G.S.}}\ ,\label{gapH}
\end{equation}
where the mean-field ground state energy is given as:
\begin{subequations}
\begin{align}
& E^{\text{mf}}_{\text{G.S.}}
=\sum_{{\bf k}}\left\{ \frac{3}{2}(\omega_1({\bf k}) 
+\omega_2({\bf k}))-3(J_2-\mu)-\mu(\overline{s}^2 -1)\right\}\ ,\\
& (\omega_1({\bf k})\ ,\ \omega_2({\bf k}))=
(\omega_{(+,-)}({\bf k})\ ,\ \omega_{(+,+)}({\bf k}))\ ,
\label{omegapm}\\
& \omega_{(\pm ,\pm)}
= \pm\frac{1}{2} \Bigl[ a^2-c^2-d^2+e^2 \pm \bigl\{ 
\left(-a^2+c^2+d^2-e^2\right)^2\nonumber\\
& \ \ +4 (a-d) (c-e) \left(-4 b^2+a c+c d+a e+d   e \right) 
\bigr\}^{\frac{1}{2}} \Bigr]^{\frac{1}{2}} \ .\label{pmpm}
\end{align}
\end{subequations}
In eq. (\ref{pmpm}), the order of signs $\pm$ coincides on both sides.  
Since $\omega_{1,2} \geq 0$, condensation of the triplets $p$ and $q$ 
occurs when the equality holds at some ${\bf k}$.  
Otherwise, there is no condensation, and 
$\langle p^{\prime\alpha}_{{\bf k}} \rangle
=\langle q_{{\bf k}}^{\prime\alpha} \rangle=0$. 
Therefore, there exist rotational symmetry and no magnetic order.  
In this case, $\omega_{1,2}$ are the excitation energy of triplets. 

\begin{figure}[ht]
\begin{center}
\includegraphics[scale=0.5]{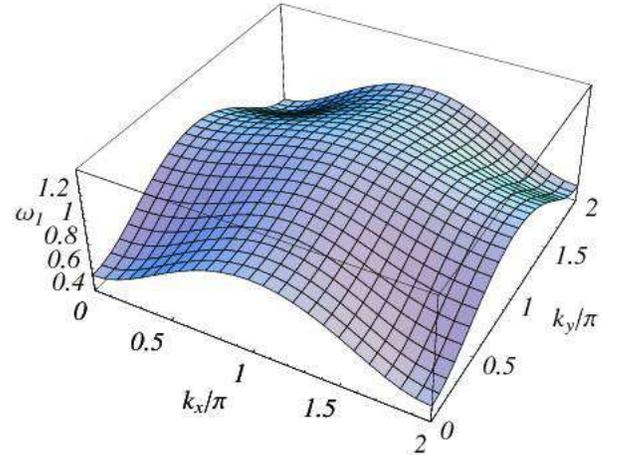}
\caption{The dispersion relation of the excitation energy 
of the triplet $p^\prime$ in (\ref{gapH}), which has the lower energy 
of the two triplets, for the parameters $\lambda=0.3,\ J_1=-0.8,\ J_2=1$}
\label{omega1}
\end{center}
\end{figure}

We looked for the solutions $(\mu,\overline{s})$ to the set of equations 
(\ref{eqn:variational}) numerically.  
For example, we found $(\mu,\overline{s})=(-0.09,0.96)$ 
for the set of parameters $\lambda=0.3,\ J_1/J_2=-0.8$. 
The dispersion relation of the excitation energy 
$\omega_1({\bf k})=\omega_{(+,-)}({\bf k})$ is shown 
in FIG.~\ref{omega1}.  

If the excitation becomes soft $\omega=0$ at some ${\bf k}$, 
the system is in a magnetically ordered phase.  
From the known results\cite{Shannon-04,Shannon-06}, 
we expect that ordered phase appears for $\lambda$ sufficiently close 
to 1.  To determine the phase boundary between the paramagnetic phase 
and magnetically ordered ones, we searched the $(\lambda,J_1/J_2)$ plane 
for the points where the mean-field gap vanishes.  
Unfortunately we found that the gap did not close in the relevant 
parameter region $0<\lambda<1,\ -2<J_1/J_2 <0$, and that 
the disordered singlet phase persisted; the gap vanished only 
for larger $\lambda(>1)$. 
This unacceptable result may be attributed to the fact 
that the bond-operator mean-field theory probably overestimates 
the stability of the plaquette phase.

\subsection{Second order perturbation}
\label{sec:2per}
In this section, we compute the energy gap of triplets 
$\ket{1,0;1}$, $\ket{0,1;1}$ and the quintet $\ket{1,1;2}$ 
by the second order perturbation theory in the distortion 
parameter $\lambda$.  The naive expansion in $\lambda$ is 
ill-behaved in the vicinity of the point $J_{1}/J_{2}=-2$ 
and we have to use another perturbation scheme for that region.  
\subsubsection{The excitation energies of triplets}
Let us consider the states where there exists only one triplet 
and all the other plaquettes are in the singlet $\ket{0,0;0}$ state. 
If the coupling constant of inter-plaquette interaction $\lambda$=0, 
these states are $N_{\text{p}}$-fold degenerate, 
where $N_{\text{p}}$ is the number of plaquettes.  
For finite $\lambda$, the second order perturbation 
induces hopping of the triplet to nearest or next nearest neighbors 
and lifts the degeneracy. 

Rotational symmetry forbids the hopping 
which changes the spin label $i(=x,y,z)$ or the magnetic quantum
number.  On the other hands, the transitions between two different triplets 
$p_i$ and $q_i$ of the same label $i$ occur. 
For example, the hopping amplitude of $p_i(q_i)$ to 
the nearest-neighbor plaquette is given by
\begin{eqnarray}
\begin{split}
-\frac{\lambda}{4}J_{2}-\frac{\lambda^{2}}{8}J_{2}
\end{split}
\end{eqnarray}
The degeneracy is partially resolved by the hopping of $p_i(q_i)$. 
The transition between $p_i$ and $q_i$ will be considered later.  
In the second-order perturbation, the processes  
that the triplet returns to the original site is also allowed.  
Including this effect, the energy change of $p_i({\bf k})$-particle 
is given by $a({\bf k})$ in (\ref{AP2:p}).  
Similarly, that of $q_{i}({\bf k})$ is given by $b({\bf k})$ 
in (\ref{AP2:q}).  
 
\begin{figure}[H]
\begin{center}
\includegraphics[scale=0.5]{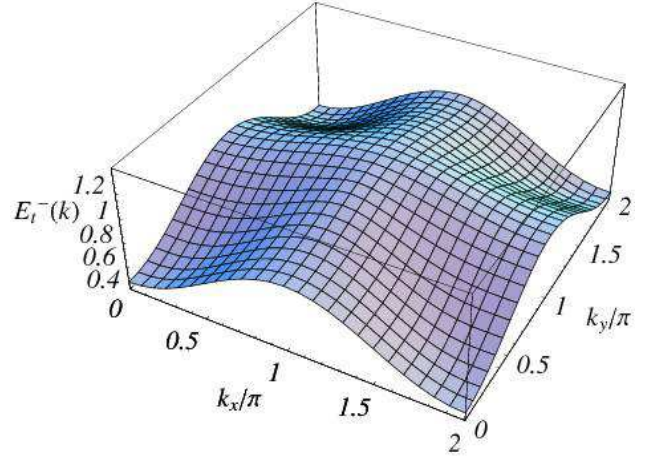}
\caption{The dispersion relation of the excitation energy 
$E_t^{-}({\bf k})$ of triplet at $\lambda=0.3,\ J_1=-0.8,\ J_2=1$.
\label{bunsan1}}
\end{center}
\end{figure}

Next, we consider the transition between $p_i({\bf k})$ and 
$q_i({\bf k})$. The transition amplitude is given by
\begin{eqnarray}
\begin{split}
c({\bf k})\equiv 
\left(\frac{J_1}{2}\lambda 
+\frac{J_{1}^{2}}{4J_2}\lambda^2 \right) 
(\cos k_{x}-\cos k_{y}) \; .
\end{split}
\end{eqnarray}
Therefore, for each $i=(x,y,z)$, eigenstates ${\bf t_+},{\bf t_-}$ satisfies
\begin{eqnarray}
\left(
\begin{array}{cc}
a({\bf k}) & c({\bf k})\\
c({\bf k}) & b({\bf k})\\
\end{array}
\right)
\left(
\begin{array}{cc}
t^{(1)}_{\pm}\\
t^{(2)}_{\pm}\\
\end{array}
\right)
=E({\bf k})
\left(
\begin{array}{cc}
t^{(1)}_{\pm}\\
t^{(2)}_{\pm}\\
\end{array}
\right)\label{acb}
\end{eqnarray}
The expressions of $a({\bf k}),b({\bf k}),c({\bf k})$ are given 
in Appendix~\ref{sec:AP2}.
After this procedure, the degeneracy with respect both to the position and  
to the species $p_i$ and $q_i$ is resolved. 
There also exists an energy shift in the ground state.  
Taking all these into account, we obtain the energy of the triplets:
\begin{eqnarray}
\begin{split}
& E_t^{\pm}({\bf k})=\\
& \frac{1}{2} \left(a({\bf k})
+b({\bf k})\pm\sqrt{(a({\bf k})
-b({\bf k}))^2+4 c({\bf k})^2}\right)-\Delta E_{\text{s}}\ ,\label{Etpm}
\end{split}
\end{eqnarray}
where $\Delta E_{\text{s}}$ denotes the energy shift of the bare ground state 
where all plaquettes are occupied by the singlet $\ket{0,0;0}$ 
and is given by eq.(\ref{AP2:Es}). 
The dispersion relation of the lower branch $E_t^{-}$ is shown 
in FIG.~\ref{bunsan1}
The lower branch $E_{t}^{-}$ takes its minimum at the $\Gamma$-point 
${\bf k}=0$, and $E_{t}^{-}({\bf k}=0)$ gives spin gap 
$\Delta_{t}$.  
The second order expression of $\Delta_{t}$ is given in (\ref{AP2:tmin}).  
The expression tells us that $\Delta_t$ has a pole at $J_1 =-2J_2$  
and that the standard perturbation breaks down near the pole.  
To remedy this, we introduce another perturbation parameter 
$\delta=J_{1}-(-2J_{2})$ and carry out a double expansion in both 
$\lambda$ and $\delta$.  Then, we obtain the energy gap 
in a modified method $E_{t,\text{mod}}^{-}(0)$ given 
in eq.(\ref{AP2:tmind}). 
This improved energy gap is expressed to give a better approximation 
around $J_{1}=-2J_{2}$
\subsubsection{excitation energy of quintet}
\begin{figure}[H]
\begin{center}
\includegraphics[scale=0.5]{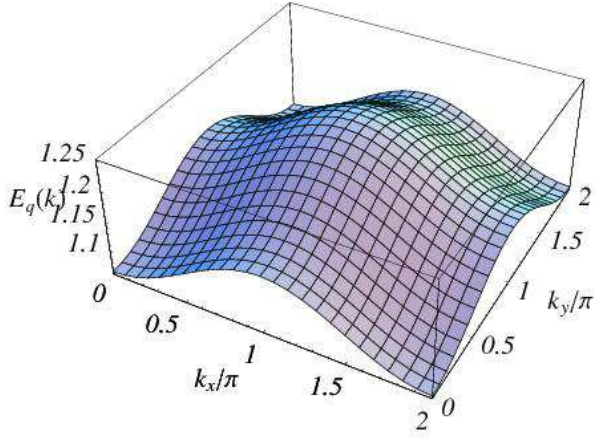}
\caption{The dispersion relation of the excitation energy 
$E_q({\bf k})$ 
of quintet at $\lambda=0.3,\ J_1=-0.8,\ J_2=1$.\label{bunsanq}}
\end{center}
\end{figure}

Next, we consider states containing only one quintet 
in a background of the singlet plaquettes.  
As before, the degeneracy with respect to the position of 
the quintet plaquette is resolved by hopping.  
Up to the second order in $\lambda$, the hopping to nearest neighbor 
is given by
\begin{eqnarray}
\lambda^2\frac{J_{1}^{2}+J_{2}^{2}}{8J_{1}}
\end{eqnarray}
and the hopping to next nearest neighbor does not occur.  
Taking into account the processes that the quintet returns to 
the original site and the energy shift of the ground state, 
the excitation energy of quintet is given by $E_q({\bf k})$ 
in (\ref{AP2:quin}). The dispersion relation is shown in FIG.~\ref{bunsanq}
Since the quintet dispersion $E_q({\bf k})$ takes its minimum at 
${\bf k}={\bf 0}$,  
the quintet gap is given by
\begin{equation}
\begin{split}
\Delta_{q}& \equiv 
E_q({\bf k}={\bf 0}) \\
&=J_{1}+2J_{2}-\frac{\lambda ^{2}(6J_{1}^{3}-23J_{1}J_{2}
+10J_{2}^{3})}{8J_{1}(2J_{1}-J_{2})}\ .
\end{split}
\label{quingap}
\end{equation}
We note that there is the pole at $J_1 =0$ and the approximation 
becomes poor for $J_{1}\approx 0$.

\section{Ground State Phases}
\label{sec:phase}
\begin{figure}[ht]
\begin{center}
\includegraphics[scale=0.6, angle=270]{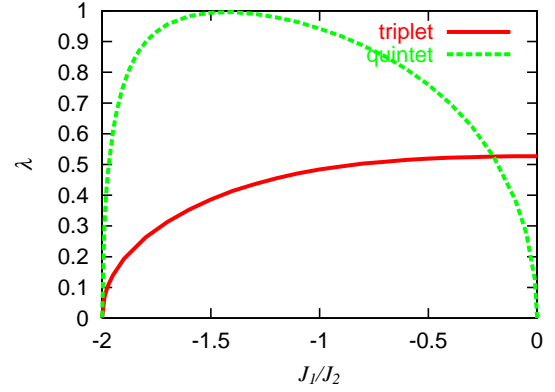}
\caption{The value of $\lambda$ at which the energy gaps 
$\Delta_{t}$ and $\Delta_{q}$ close. 
The energy gap of the triplets $\Delta_{t}$ is given in (\ref{AP2:tmin}), 
which is not reliable near $J_1/J_2=-2$ because of the pole there, 
and that of the quintet $\Delta_{q}$ is in (\ref{quingap}), 
which is not reliable near $J_1/J_2=0$.%
\label{fig:qt}}
\end{center}
\end{figure}

\begin{figure}[ht]
\begin{center}
\includegraphics[scale=1.2]{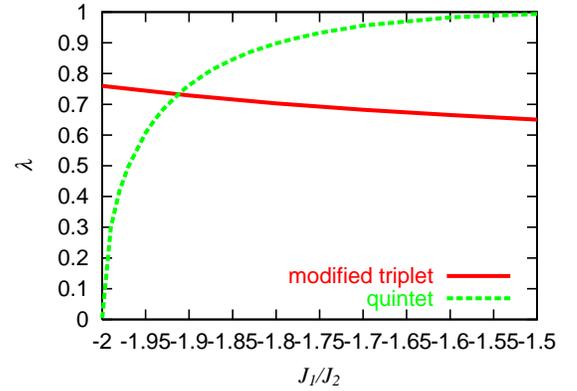}
\caption{Plot of the value of $\lambda$ 
when the energy gaps are equal to 0.  
We use $\Delta_{t,\text{mod}}$ (eq.(\ref{AP2:tmind})) for the triplets, 
which is reliable even in the vicinity of $J_1/J_2=-2$. 
The curve for the quintet is the same as in FIG.~\ref{fig:qt}.  
We only plot the region $-2<J_1/J_2<-1.5$.%
\label{fig:qtmod}}
\end{center}
\end{figure}

\begin{figure}[ht]
\begin{center}
\includegraphics[scale=0.4]{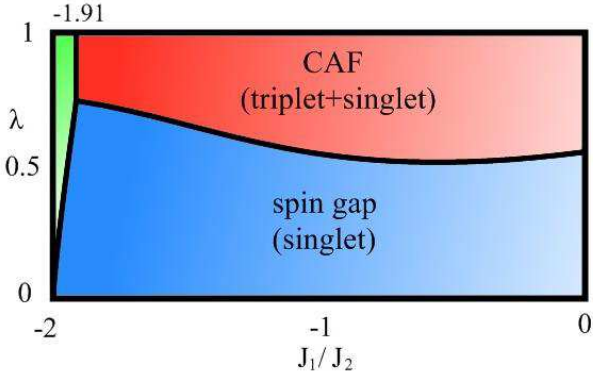}
\caption{The schematic phase diagram of the ground state 
determined by the particle whose excitation gap closes first. 
In the green region, the quintet and the singlet condense, 
in the red do the triplet and the singlet, and in the blue does the singlet. 
In the region marked by blue, the energy gap exists. The phase shown 
by red may be considered as collinear antiferromagnetic (CAF) state. 
The nature of the green phase, where the quintet condensation occurs, 
is closely investigated by using an effective Hamiltonian 
$H_{\text{qu}}$ (eq.(\ref{Hqumean})).%
\label{fig:souzuG}}
\end{center}
\end{figure}

If the inter-plaquette coupling $\lambda$ is increased, 
one of the energy gaps of the triplets ((\ref{AP2:tmin}) 
and (\ref{AP2:tmind})) and the quintet (\ref{AP2:quin}) becomes 
$0$ at a certain critical value of $\lambda$.  When it happens, 
the corresponding particle condenses and a phase transition occurs 
from the gapped spin-singlet phase to superfluid phases with magnetic 
long-range order.  
Therefore, we can classify the phases according to what kind of 
particles condense and what kind of magnetic orders is stabilized 
by a given set of interactions among them.  
In FIGs.~\ref{fig:qt} and \ref{fig:qtmod}, 
we plot the value of $\lambda$ at which the smallest energy gap becomes 0.
If we assume that no further condensation occurs in 
the other kinds of particles once the triplets or the quintet condenses, 
the phase diagram FIG.~\ref{fig:souzuG} is obtained. 
When we mapped out the phase diagram FIG.~\ref{fig:souzuG}, 
we have used two different expressions (\ref{AP2:tmind}) 
and (\ref{AP2:tmin}) for the energy gap of the lowest 
triplet in the vicinity of $J_1/J_2=-2$ and   
away from it ($J_1/J_2\sim 0$), respectively.  
We have also neglected the quintet around $J_1/J_2=0$ since 
the collapse of the quintet gap there (see FIG.~\ref{fig:qt}) can be  
attributed to the existence of a pole and is just an artifact of 
the perturbative approximation. 
Note that the phase boundary between the two regions covered by  
eq.(\ref{AP2:tmin}) and eq.(\ref{AP2:tmind}) is only schematic. 

Now let us discuss the nature of the ordered phases appearing after 
the condensation.  In the region shown as ``CAF'' (highlighted in red) 
in FIG.~\ref{fig:souzuG}, condensation occurs to the singlet and 
the triplets. Then, we may expect:
\begin{eqnarray}
|\langle s\rangle |\neq 0,\ |\langle {\bf p}({\bf k}={\bf 0})\rangle 
|^{2}\neq 0,\ 
|\langle {\bf q}({\bf k}={\bf 0})\rangle |^{2}\neq 0\ ,
\end{eqnarray}
which, combined with (\ref{bt4}), implies 
\begin{subequations}
\begin{align}
& \langle {\bf S}_{1}\rangle =-\langle {\bf S}_{3}\rangle \neq 0\ ,\\
& \langle {\bf S}_{2}\rangle =-\langle {\bf S}_{4}\rangle \neq 0\ ,
\end{align}
\end{subequations}
provided that $\epsilon^{\alpha\beta\gamma}\langle p^\dagger _\beta\rangle 
\langle p_\gamma\rangle=0$ and $\epsilon^{\alpha\beta\gamma}\langle 
q^\dagger _\beta\rangle \langle q_\gamma\rangle=0$. 
Note that all the plaquettes are in the same state, 
since the energy of the triplet takes its minimum at the $\Gamma$-point 
${\bf k}=0$ (see FIG.~\ref{bunsan1}). 
When the combination $({\bf p}+{\bf q})$ of the two bosons condenses,  
the relation $\langle{\bf S}_{1}\rangle=\langle{\bf S}_{2}\rangle$ 
holds and the ground state has the transversely aligned 
(i.e. $(0,\pi)$) collinear antiferromagnetic order.  
In the case where $({\bf p}-{\bf q})$ condenses, 
on the other hand, we have $\VEV{{\bf S}_{1}}=-\VEV{{\bf S}_{2}}$ 
instead and the system is in the collinear antiferromagnetic ground 
state in the longitudinal ($(\pi,0)$) direction.  
This is consistent with the known results\cite{Shannon-04}.  

Now we move on to a more interesting case.  
In the green region in FIG.~\ref{fig:souzuG}, frustration is strong 
($J_2 \approx -2 J_1$) and nontrivial order may be expected.  
In fact, Shannon {\em et al.}\cite{Shannon-04,Shannon-06} analyzed 
the uniform $(\lambda=1)$ model by numerical exact diagonalizations 
up to clusters of 36 spins and found a spin-nematic phase 
with $d$-wave (or $\text{B}_{1}$) symmetry 
for $-2.5<J_1/J_2<-1.43\sim -1.67$.  In the state with 
the nematic order, the expectation value of the rank-1 tensor vanishes 
$\langle{\bf S}_i \rangle =0$, while we have a finite expectation 
value of the following traceless rank-2 tensor:
\begin{equation}
Q^{\alpha \beta}_{ij}\equiv \frac{S_{i}^{\alpha}S^{\beta}_{j}
+S^{\beta}_{i}S^{\alpha}_{j}}{2}-\frac{{\bf S}_{i}\cdot {\bf S}_{j}}{3}
\delta^{\alpha\beta}\ ,
\end{equation}
where ${\alpha,\beta=(x,y,z)}$ and $i,j$ label the lattice sites.

As is shown in FIG.~\ref{fig:souzuG}, 
the singlet and the quintet condense in the region of interest.  
This is analogous to the spinor Bose-Einstein condensation of 
spin-2 particles (here particles are defined not on the lattice sites 
but on the plaquettes).   
We consider a single plaquette (see FIG.~\ref{fig:plas}) and, 
as before, denote the singlet and the quintet respectively by 
$\ket{s}$ and $\ket{1,1;2,Sz}$.  
To investigate what kind of magnetic order is stabilized in 
the condensate, let us introduce the following mean-field ansatz for 
the ground state:
\begin{multline}
|\{\theta_{\bolr}\},\{\psi_{\bolr}\}\rangle = \\
\bigotimes_{\bolr\in\text{plaq}}
\left\{
\cos\theta_{\bolr}|\text{s}\rangle_{\bolr}
+ \sin\theta_{\bolr}\sum_{S_{z}=-2}^{2}\psi_{\bolr}(S_{z})
\ket{1,1;2,S_{z}}_{\bolr} 
\right\}  \; ,
\end{multline}
where the product is over all plaquettes and the complex numbers 
$\psi_{\bolr}(S_{z})$ satisfy $\sum_{S_{z}}|\psi_{\bolr}(S_{z})|^{2}=1$.   
Then, since the rank-1 tensor can not give rise to transitions 
between the spin-0 states and the spin-2 ones 
by the Wigner-Eckart theorem\cite{JJWE}, we have 
$\bra{s}S^{\alpha}_i\ket{s}=\bra{s}Q_{ij}^{\alpha\beta}\ket{s}=0$, 
$\bra{s}S^{\alpha}_i\ket{1,1;2,Sz}=0$ and consequently 
$\langle\{\theta_{\bolr}\},\{\psi_{\bolr}\}|{\bf S}_{i}
|\{\theta_{\bolr}\},\{\psi_{\bolr}\}\rangle={\bf 0}$. 
If we introduce the cyclic 
operator $C$ which translates the state as 
$1\rightarrow 2\rightarrow 3 \rightarrow 4\rightarrow 1$, 
we obtain $C\ket{s}=-\ket{s},\ C\ket{1,1;2,Sz}=\ket{1,1;2,Sz}$ from (\ref{ss1}) and (\ref{qq}).  
Therefore, the spin-nematic tensor $Q^{\alpha\beta}_{ij}$ defined 
on the bond $(i,j)$ satisfies 
$\bra{s}Q_{12}^{\alpha\beta}\ket{1,1;2,Sz}
=\bra{s}C^{\dagger}C Q_{12}^{\alpha\beta} C^{\dagger}C\ket{1,1;2,Sz}
=-\bra{s} Q_{23}^{\alpha\beta}\ket{1,1;2,Sz}=\bra{s}Q_{34}^{\alpha\beta}\ket{1,1;2,Sz}
=-\bra{s} Q_{41}^{\alpha\beta}\ket{1,1;2,Sz}$. 
This implies that the spinor condensate 
$|\{\theta_{\bolr}\},\{\psi_{\bolr}\}\rangle$ of our quintet boson has 
the same ($d$-wave) symmetry as the spin-nematic state discussed in
Ref.\onlinecite{Shannon-06}.  

However, this is not the end of the story.  
Since the local spin operator with $S \geq 1$ assumes several different 
states (e.g. {\em polarized}, {\em nematic}, etc.) and it is not 
obvious if 
$\langle\{\theta_{\bolr}\},\{\psi_{\bolr}\}|Q_{ij}^{\alpha\beta}
|\{\theta_{\bolr}\},\{\psi_{\bolr}\}\rangle\neq 0$ or not 
for our $J_{1}$-$J_{2}$ model.   To determine the actual value 
of $\langle\{\theta_{\bolr}\},\{\psi_{\bolr}\}|Q_{ij}^{\alpha\beta}
|\{\theta_{\bolr}\},\{\psi_{\bolr}\}\rangle$, we need the explicit 
mean-field solution for a given set of $(J_1,J_2,\lambda)$. 
Since we are considering the situation where the gap between 
the singlet ground state and the quintet excitation is vanishingly 
small, it would be legitimate to keep only the singlet 
$|0,0,0\rangle$ and the quintet for each plaquette to write down 
the effective Hamiltonian.    

\begin{figure}[ht]
\begin{center}
\includegraphics[scale=0.5]{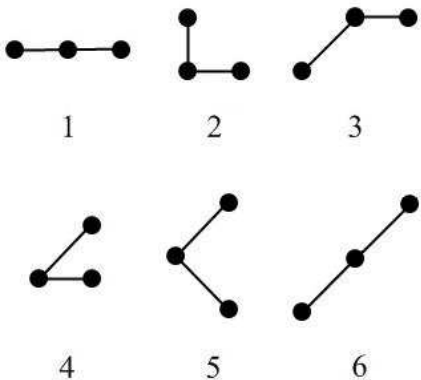}
\caption{Clusters involved in the 3-points interaction in (\ref{Hqumean}). 
The plaquette corresponding to $i^{\prime\prime}$ is always located 
on the center of the clusters. 
We identify all clusters obtained from a given one by 
rotation by $\pi /2,\pi,3\pi /2$ and reflection.   
\label{fig:3point}}
\end{center}
\end{figure}

The form of the effective Hamiltonian is determined by 
using the second-order perturbation theory and it contains 
the kinetic part describing the hopping of the quintet particles and 
the magnetic part which concerns the interactions among them.  
Since within a mean-field treatment the spinor part 
$\psi_{\bolr}(S_{z})$ is determined by 
the magnetic interactions, 
it suffices to consider only the magnetic part of the effective 
Hamiltonian: 
\begin{widetext}
\begin{equation}
\begin{split}
& H_{{\rm qu}}=\sum_{\langle i,j \rangle} \left\{
J_{{\rm qu1}}(\widetilde{{\bf S}} ^q_i {\cdot}\widetilde{{\bf S}}^q_j)
+K_{{\rm qu} 1}(\widetilde{{\bf S}} ^q_i{\cdot}\widetilde{{\bf S}}^q_j)^2 
\right\}
+\sum_{\VEV{i^{\prime},j^\prime}} \left\{
J_{\rm qu2}(\widetilde{{\bf S}}^q_{i^\prime} {\cdot} 
\widetilde{{\bf S}}^q_{j^\prime})
+K_{{\rm qu} 2}(\widetilde{{\bf S}}^q_{i^\prime}{\cdot} 
\widetilde{{\bf S}}^q_{j^\prime})^2 \right\}
+\sum_{\VEV{i^{\prime\prime},j^{\prime\prime},k^{\prime\prime}}} \\
& \left[
L_{{\rm qu}1}\left\{ (\tilde{{\bf S}} ^q_{i^{\prime\prime}} {\cdot} 
\widetilde{{\bf S}} ^q_{j^{\prime\prime}})
(\widetilde{{\bf S}}^q_{i^{\prime\prime}} 
{\cdot}\widetilde{{\bf S}} ^q_{k^{\prime\prime}})+
(\widetilde{{\bf S}}^q_{i^{\prime\prime}} {\cdot} 
\widetilde{{\bf S}}^q_{k^{\prime\prime}})
(\widetilde{{\bf S}}^q_{i^{\prime\prime}} 
{\cdot}\widetilde{{\bf S}}^q_{j^{\prime\prime}})\right\}
+L_{{\rm qu}2}\left\{(\widetilde{{\bf S}}^q_{i^{\prime\prime}}
{\times} \widetilde{{\bf S}}^q_{j^{\prime\prime}}){\cdot} 
(\widetilde{{\bf S}}^q_{i^{\prime\prime}}{\times} 
\widetilde{{\bf S}}^q_{k^{\prime\prime}}) 
+(\widetilde{{\bf S}}^q_{i^{\prime\prime}}{\times} 
\widetilde{{\bf S}}^q_{k^{\prime\prime}}){\cdot} 
(\widetilde{{\bf S}}^q_{i^{\prime\prime}}{\times} 
\widetilde{{\bf S}}^q_{j^{\prime\prime}})\right\} \right],
\end{split}
\label{Hqumean}
\end{equation}
\end{widetext}
where $\widetilde{\bf S}^q$ denotes the $S=2$ spin operator, 
and the symbols $\VEV{i,j}$ and $\VEV{i^\prime,j^\prime}$ mean 
the nearest-neighbor- and the next-nearest-neighbor pairs, 
respectively.    
For different types of three-plaquette clusters 
$\VEV{i^{\prime\prime},j^{\prime\prime},k^{\prime\prime}}$, 
we assign different three-body (i.e. three-plaquette) interactions 
$L^{(n)}_{\text{qu}1,2}$ ($n=1,\dots,6$) in (\ref{Hqumean}).  
The correspondence between six types of clusters and the strength of 
the three-plaquette interaction $L^{(n)}_{\text{qu}1,2}$ 
is shown in FIG.~\ref{fig:3point}.  
The full expressions of $J_{\text{qu}1,2}$, $K_{\text{qu}1,2}$ 
and $L^{(n)}_{\text{qu}1,2}$ are given in Appendix.~\ref{sec:AP2}. 
Note that our effective Hamiltonian in its full form contains 
the kinetic term and charge interactions as well as magnetic ones 
$H_{{\rm qu}}$.  
In this sense, our effective model is a generalization of the 
Bose Hubbard Hamiltonian for $F=2$ cold atoms in optical 
lattices\cite{Barnett,Zhou-Semenoff,Lewenstein-review} 
and the determination of the full phase diagram and the identification 
of various phases found in systems of cold atoms in our magnetic system 
would be interesting in its own right.  

\begin{figure}[ht]
\begin{center}
\includegraphics[scale=0.45]{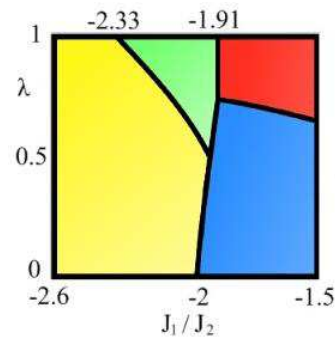}
\caption{The schematic phase diagram obtained 
in a similar manner to in FIG.~\ref{fig:souzuG}.  
We zoom up the region around $J_{1}/J_{2}=-2$ in FIG.~\ref{fig:souzuG}. 
In the two regions on the left (green and yellow), the quintet and 
the singlet condense and we determined the resulting magnetic orders 
by a mean-field approximation to the magnetic Hamiltonian $H_{\text{qu}}$.   
In the green region, the quintet and the singlet condense, and 
the spin-nematic phase appears. In the red, on the other hand, 
conventional ferromagnetic order is stabilized.  
The red and the blue region are the same as FIG.~\ref{fig:souzuG}.
\label{fig:nf}}
\end{center}
\end{figure}

We investigate this Hamiltonian by means of a mean-field theory by assuming 
an $\bolr$-independent uniform $\{\theta,\psi\}$, for simplicity. 
Since the parametrization of the spin-2 states is cumbersome, 
we adopt the method used by Bacry\cite{Bacry} and Barnett 
{\em et al.} \cite{Barnett}.  
First we note that arbitrary (normalized) spin-$S$ states are parametrized 
by a set of $2S$ unit vectors except for obvious gauge redundancy.  
Using rotational symmetry, we can further reduce the number of free 
parameters needed to express arbitrary spin-2 states 
to $2{\times}4{-}3{=}5$ (see Appendix.~\ref{classify}).  
We numerically minimized the mean-field 
energy with respect to these five parameters.  
The result is shown in FIG.~\ref{fig:nf}.

At $\lambda=1$, the system is in the ferromagnetic 
state for $J_1/J_2<-2.33$ and is in the spin-nematic state 
for $-2.33\leq J_1/J_2(\leq -1.91)$.  This result slightly differs from 
the numerical results\cite{Shannon-04,Shannon-06} 
$-2.5\lesssim J_1/J_2\lesssim -1.43 \sim -1.67$.   
However, this is not surprising since our results are based on 
a mean-field treatment of the magnetic Hamiltonian $H_{\text{qu}}$ 
obtained by perturbation expansion in $\lambda$.   
Our result may be improved by taking the number of sublattice larger, 
since $J_{{\rm qu}2}>0$ and there are various 3-site interactions 
$L_{\text{qu1}}$ and $L_{\text{qu2}}$.
\section{Magnetization process}
\label{sec:magnet}
Having mapped out the phase diagram in the absence of external 
magnetic field, 
we consider next the magnetization process of the plaquette model 
by mapping the original model onto a hardcore boson model 
or an equivalent $S=1/2$ pseudo-spin model.  
Tachiki and Yamada\cite{TY} applied this method to obtain 
the magnetization curve of the spin-dimer model, 
which consists of pairs of $S=1/2$ spins. 
The coupling to the external magnetic field is incorporated into 
the Hamiltonian by adding the Zeeman term 
$g\mu_{B}{\bf h}{\cdot}\sum_{i}{\bf S}_{i}$.  
For convenience, we set $g\mu_{\text{B}}=1$ and assume that 
${\bf h}$ is pointing the $z$-direction: ${\bf h}=(0,0,h)$.  

\begin{figure}[H]
\begin{center}
\includegraphics[scale=0.64]{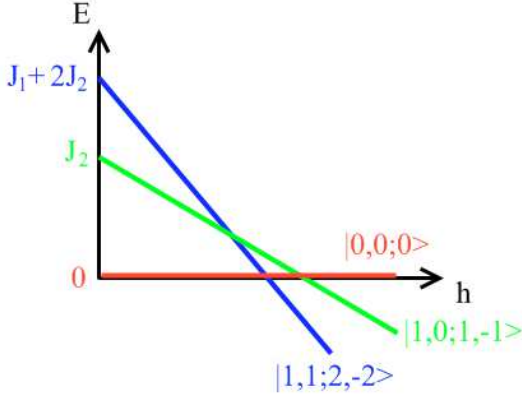}
\caption{The energy of eigenstates of a single plaquette 
as a function of magnetic field $h$.}
\label{Fig:jikas}
\end{center}
\end{figure}

Although the original treatment in Ref.~\onlinecite{TY} is for 
a coupled dimer systems, we can readily generalize the method 
to our plaquette system as follows.  
We denote the plaquette states by $|S_{a},S_{b};S,S^{z}\rangle$, 
where $S,S_{a},S_{b}$ are defined in ($\ref{Hpsingle}$).  
From (\ref{sec2:EE}), the energies of a single plaquette satisfy
\begin{eqnarray}
E(1,1,2)<2E(1,0,1),\ 2E(0,1,1)\ ,
\end{eqnarray}
for \(-2 < J_{1}/J_{2} < 0 \).  As is shown in FIG.~\ref{Fig:jikas}, 
with increasing the magnetic field, the quintet level $\ket{1,1;2,-2}$ 
comes down to $\ket{0,0;0}$ faster than the lowest triplet levels 
$\ket{1,0;1-1}$ and $\ket{0,1;1,-1}$.  
Therefore, in order to describe the low-energy physics 
in the presence of strong magnetic field ($h\sim J_{1}/2+J_{2}$), 
we may keep only the two lowest-lying states 
$\ket{0,0;0}$ and $\ket{1,1;2,-2}$ for each plaquette and 
restrict ourselves to the subspace spanned by them.   
In what follows, we regard the singlet $\ket{0,0;0}$ and 
the quintet $\ket{1,1;2,-2}$ respectively as the up- and the down state 
of a pseudo spin-1/2.   
That is, 
\begin{equation}
|0,0;0\rangle=
\begin{pmatrix}
1\\
0
\end{pmatrix}
\; , \quad 
|1,1;2,-2\rangle=
\begin{pmatrix}
0 \\
1
\end{pmatrix}  \; .
\label{eqn:def-pseudospin}
\end{equation}
Then, the resulting effective Hamiltonian is written in terms of 
the Pauli matrices ($S=1/2$ spins) defined on each 
strongly-coupled plaquette.   

Note that the approximation to treat only the subspace spanned 
by $\ket{0,0;0}$ and $\ket{1,1;2,-2}$ probably breaks down for $h\approx0$ 
where all the components ($S^{z}=-2,\dots,2$) of the quintet 
come into play.  
Also the validity of the approximation may be questionable 
for sufficiently large $\lambda$ where the singlet-triplet gap may be 
much smaller than the singlet-quintet gap, since the triplet 
states $|1,0;1\rangle$ and $|0,1;1\rangle$ are important there  
(see FIG.~\ref{fig:souzuG}).  

If we simply project the original $S=1/2$ Hamiltonian to 
the restricted subspace as in (\ref{psgS}), 
no spin-flipping term (or, hopping term, in terms of hardcore bosons) 
appears.  This is because the projection is equivalent to  
the ordinary first-order perturbation theory and no transition 
between the singlet and the quintet occurs in the first-order
processes.   
Therefore, we need take into account the second-order processes 
to obtain the meaningful effective Hamiltonian.  
The amplitude that a quintet state (spin `down') $\ket{1,1;2,-2}$ 
hops to the adjacent plaquette is given by
\begin{equation}
t\equiv \lambda^{2}\frac{J_1^2+J_2^2}{8 J_1}\ .
\end{equation}
The hopping to the next nearest-neighbor does not occur at this order 
of approximation.  
The energy gap between the state where there exists only one 
static `down' spin ($\ket{1,1;2,-2}$) in a background of the `up' spins 
(singlet $\ket{0,0;0}$ plaquettes) 
and the one where all plaquettes are `up' is given by $-\mu$ 
in (\ref{AP2:mu}). The interaction between the two adjacent 
`up' spins ($\ket{1,1;2,-2}$) is given by $J_{{\rm eff}1}$ in (\ref{AP2:Je1}) 
and that between the next-nearest-neighbor pair is given by 
$J_{{\rm eff}2}$ in (\ref{AP2:Je2}).  
We note that this approximation becomes poor near the pole  
of $J_{{\rm eff}1,2}$ and $t$ at $J_1/J_2=0$. 
On top of them, we have several three-`site' processes and 
putting them all together, we obtain the effective Hamiltonian: 
\begin{widetext}
\begin{equation}
\begin{split}
H_{\text{eff}}& =
(J_{1}+2J_{2}-\mu-2h)\sum_{i}\sigma^{-}_i\sigma^{+}_i
+\sum_{\VEV{i,j}}\left\{t(\sigma^{+}_i \sigma^-_j+\sigma^{-}_i \sigma^+_j)
+J_{\text{eff}1}(\sigma^{-}_i\sigma^{+}_i)(\sigma^{-}_j\sigma^{+}_j)\right\}\\
&  +\sum_{\VEV{i^{\prime},j^{\prime}}}J_{\text{eff}2}
(\sigma^{-}_{i^{\prime}}\sigma^{+}_{i^{\prime}})
(\sigma^{-}_{j^{\prime}}\sigma^{+}_{j^{\prime}}) 
+\sum_{\VEV{i^{\prime\prime},j^{\prime\prime},k^{\prime\prime}}}
L_{\text{eff}}(\sigma^{+}_{i^{\prime\prime}}\sigma^{-}_{i^{\prime\prime}})
(\sigma^{-}_{j^{\prime\prime}}\sigma^{+}_{j^{\prime\prime}})
(\sigma^{-}_{k^{\prime\prime}}\sigma^{+}_{k^{\prime\prime}})\ ,
\end{split}
\label{Heff}
\end{equation}
\end{widetext}
where ${\bf \sigma}$s denote the Pauli matrices and 
$\sigma^{+}=\frac{1}{2}(\sigma^x+i\sigma^y),\ 
\sigma^{-}=\frac{1}{2}(\sigma^x-i\sigma^y)$. 
The symbols $\VEV{i,j}$ and $\VEV{i^\prime,j^\prime}$ mean that 
the summation is taken over the nearest-neighbor- and 
the next-nearest-neighbor plaquettes, respectively.  
As in section~\ref{sec:phase}, there are six types of 
$L_{{\rm eff}}$ for different bond configurations 
$\VEV{i^{\prime\prime},j^{\prime\prime},k^{\prime\prime}}$ 
(see FIG.~\ref{fig:3point}).  
We label the different three-plaquette interactions by $L^{(n)}_{\text{eff}}$ 
$(n=1\dots 6)$ and the corresponding bond-configurations are shown  
in FIG.~\ref{fig:3point}. The concrete expressions of $J_{\text{eff}}$ 
and $L_{\text{eff}}$ are given in Appendix.~\ref{sec:AP2}.  
We note that the transverse components $\sigma^{+}$ and $\sigma^{-}$ 
can be translated to the creation- $a$ and the annihilation 
$a^{\dagger}$ operator of a hardcore boson, respectively.   

\begin{figure}[H]
\begin{center}
\includegraphics[scale=0.6]{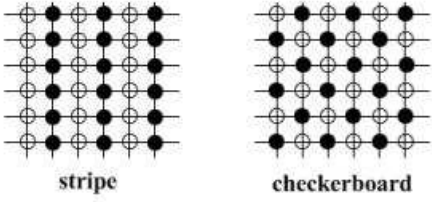}
\caption{
Two-sublattice structures assumed in the calculation: 
(i) striped- (left) and (ii) checkerboard (right) case. 
Circles (whether filled or open) denote the strongly-coupled 
plaquettes shown by thick lines in FIG.~\ref{fig:plafull}.%
\label{Fig:sublattice}}
\end{center}
\end{figure}

We analyze the Hamiltonian (\ref{Heff}) within a mean-field 
approximation.  
Since $J_{{\rm eff}1,2}$, which have the first order contributions in  
$\lambda$, are dominant for small $\lambda$, 
we may assume two different two-sublattice 
structures: (i) ``checkerboard'' and (ii) ``stripe'' shown in FIG.~\ref{Fig:sublattice} in the calculation .
By using the relations 
\begin{subequations}
\begin{align}
\sigma^{-}\sigma^{+}&= \frac{1}{2}(1-\sigma^z)\ ,\\
\sigma^{-}_i\sigma^{+}_j+\sigma^{+}_i\sigma^{-}_j &= 
\frac{1}{2}(\sigma^x_i\sigma^x_j+\sigma^y_i\sigma^y_j) \; ,
\end{align}
\end{subequations}
we can rewrite (\ref{Heff}) in terms of $\sigma^{i}(i=x,y,z)$.  
Since we are interested in the ground state energy at $T=0$,  
we can simply replace operators in (\ref{Heff}) by their expectation values 
on each site, e.g. 
$\sum_{\VEV{i,j}}\sigma^z_i \sigma^z_j \rightarrow 
\sum_{\VEV{i,j}}\VEV{\sigma^z}\VEV{\sigma^{z\prime}}$ for 
the ``checkerboard'' case.  
For convenience, we introduce the following two-component vector:
\begin{equation}
\boldsymbol{\tau}\equiv
\left(
\begin{array}{c}
\VEV{\sigma^{x}}\\
\VEV{\sigma^{y}}
\end{array}
\right)\ .
\end{equation}
Since there is rotational symmetry in the $x$-$y$ plane, 
the mean-field energy is parametrized by 
$\VEV{\sigma^z}$, $\VEV{\sigma^{z\prime}}$, 
$\tau\equiv|\boldsymbol{\tau}|$, 
$\tau^{\prime}\equiv |\boldsymbol{\tau}|$ 
and the angle $\phi$ between $\boldsymbol{\tau}$ and 
$\boldsymbol{\tau}^\prime$.  
The Hamiltonian (\ref{Heff}) reduces to
\begin{widetext}
\begin{equation}
\begin{split}
E_{\text{eff}}=& N_{\text{p}}\left[ 
\left(-\frac{\left(J_1+2J_2-\mu -2h
   \right)}{2}-\xi \right)\frac{(\VEV{\sigma^z}+\VEV{\sigma^{z\prime}})}{2} 
+\alpha_1 \VEV{\sigma^z} \VEV{\sigma^{z\prime}}
+ \alpha_2\frac{\left(\VEV{\sigma^z}^2+\VEV{\sigma^{z\prime}}^{2}\right)}{2}
\right.\\
& \left.+\left( \beta_1
   \tau  \tau^{\prime}\cos \phi+ \beta_2
\frac{\left(\tau ^2+\tau^{\prime2}\right)}{2}   \right)
+\gamma_1\frac{\left(\VEV{\sigma^z} \VEV{\sigma^{z\prime}}^{2}
+\VEV{\sigma^z}^2 \VEV{\sigma^{z\prime}}\right)}{2} 
 + \gamma_2\frac{\left(\VEV{\sigma^z}^3+\VEV{\sigma^{z\prime}}^{3}\right)}{2}
\right]
\end{split}
\label{Eeff}
\end{equation}
\end{widetext}
where $N_{{\rm p}}$ denotes the total number of plaquettes and 
$\alpha,\beta,\gamma,\xi$ are given in Appendix~\ref{sec:AP2} 
both for the case of ``checkerboard'' and for the ``striped'' case.  
Correspondingly, the total magnetization is given simply as 
\begin{equation}
M = \frac{1}{2N_{p}}\sum_{i\in \text{plaq}} (1-\sigma^{z}_{i}) \; .
\end{equation}
In both cases, $\beta_{1,2}<0$ and $E_{\rm eff}$ is minimized  
for $\phi=0$.  Since any spin-1/2 states satisfy the following 
relation among the expectation values (see eq.~(\ref{sigmaI}))
\begin{equation}
\VEV{\sigma^z}^2+\tau^2=1\ ,
\end{equation}
the transverse magnetization $\tau$ can be expressed in terms of 
the longitudinal one $\VEV{\sigma^z}$. Hence, there remain 
two variational parameters $\VEV{\sigma^z}$ and $\VEV{\sigma^{z\prime}}$ 
in $E_{\text{eff}}$.  
From the definition (\ref{eqn:def-pseudospin}), the expectation values 
$\VEV{\sigma^{z}}=1$ and $\VEV{\sigma^{z}}=-1$ respectively correspond 
to the singlet state and the fully polarized (or, saturated) state.  

The critical field $h=H_{\text{c1}}$ 
which marks the onset of magnetization is given by 
$(\partial E_{{\rm eff}}/\partial \VEV{\sigma^z})_{\VEV{\sigma^z}=1}=0$ 
after substituting $\VEV{\sigma^{z\prime}}=\VEV{\sigma^z}$, i.e.
\begin{equation}
2H_{c1}=J_1+2J_2-\mu+4t\ .\label{tatiagari}
\end{equation}
The right-hand side is exactly the same as (\ref{quingap}). 

Once spin-gap closes at $h=H_{\text{c1}}$, the quintet particle 
$\ket{1,1;2,-2}$ condenses, i.e. 
$\VEV{\sigma^z}\neq 1$, $\tau\neq 0$. 
If $\VEV{\sigma^z}\neq\pm1$, $\tau\neq0$ and there exists 
a finite expectation value of $\VEV{\sigma^-}$. 
In the hardcore boson language discussed below (\ref{Heff}), $\sigma^-$ 
can be viewed as the boson annihilation operator $a$ and 
its finite expectation value $\VEV{a}\neq 0$ implies that 
Bose-Einstein condensation of the quintet particle occurs.  
In particular, if $|\VEV{\sigma^z}|\neq|\VEV{\sigma^{z\prime}}|$ and $\tau\neq\tau^\prime$ in BEC phase, 
the state is in the so-called ``supersolid'' phase\cite{SS-old}.  
For convenience, we shall call the BEC phase satisfying 
$\VEV{\sigma^z}=\VEV{\sigma^{z\prime}}$ a {\em normal BEC}. 

It should be noted that even when $\tau\neq0$,  
the transverse magnetization $\VEV{S^{\pm}}$ vanishes unlike the BEC 
in the spin-dimer model\cite{Tanaka-01}.
In fact, since the creation operator $a^{\dagger}$ of the quintet 
particle can be written in terms of the original spin operators as
\begin{equation}
\begin{split}
& a^{\dagger} = \frac{1}{2}
\left( Q^{xx}_{\text{B}_1} - Q^{yy}_{\text{B}_1} \right) 
+ i\, Q^{xy}_{\text{B}_1} \\
& \phantom{a^{\dagger}} =  \frac{1}{2}\left(
S_{1}^{+}S_{2}^{+} - S_{2}^{+}S_{3}^{+} + S_{3}^{+}S_{4}^{+} 
- S_{4}^{+}S_{1}^{+}  \right)
\\
& Q^{ab}_{\text{B}_1}  \equiv  
Q^{ab}_{12} - Q^{ab}_{23} + Q^{ab}_{34} - Q^{ab}_{41} \; ,
\end{split}
\label{eqn:quintet-creation}
\end{equation}
the existence of the condensate $\VEV{\sigma^{+}}\neq 0$ 
(or $\tau \neq 0$) implies that we have a finite expectation value 
of the following spin-nematic operator:
\begin{equation}
\left\langle 
\frac{1}{2}
\left( Q^{xx}_{\text{B}_1} - Q^{yy}_{\text{B}_1} \right) 
\pm i\, Q^{xy}_{\text{B}_1} 
\right\rangle   \; .
\end{equation}

The form (\ref{eqn:quintet-creation}) of the quintet creation operator 
suggests that we should think of the plaquette quintet $|1,1;2,2\rangle$ 
as a tightly-bound magnon pair (or {\em magnon molecule}).  

\begin{figure}[H]
\begin{center}
\includegraphics[scale=0.5]{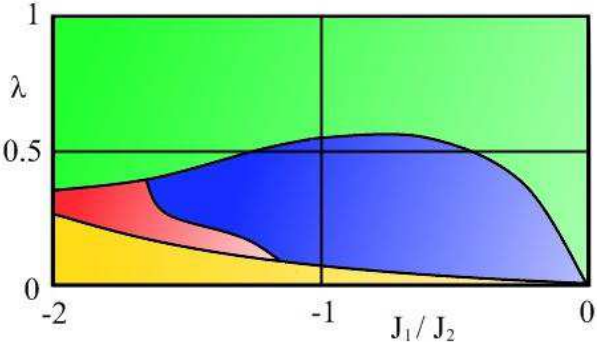}
\caption{Schematic classification of 
the magnetization curve. 
(i) In the green region, the curve is smooth and the system is 
always in the normal BEC phase. 
(ii) In the red region, the magnetization curve has a 1/2-plateau.   
Except at the plateau, the system is in the normal BEC phase. 
(iii) The region where we have additional supersolid phases 
around the 1/2-plateau is highlighted in blue. 
(iv) In the region colored by yellow magnetization jumps to saturation 
and the magnetization process is step-like. 
The concrete expression of curves is shown in FIG.~\ref{Fig:magcurve}.
The phase boundary is only schematic.%
\label{Fig:magsouzu}}
\end{center}
\end{figure}

\begin{figure}[H]
\begin{center}
\includegraphics[scale=1.2]{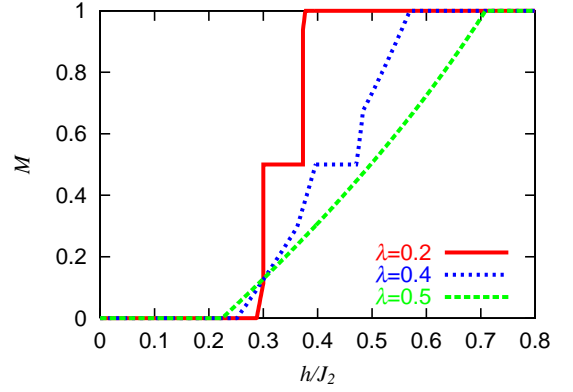}
\caption{Magnetization curve for various values of the distortion parameter 
$\lambda$.  The frustration parameters is fixed to 
$J_{1}/J_{2}=-1.4$. The colors of the curves correspond to 
those used in FIG.~\ref{Fig:magsouzu} (except for yellow). 
The blue curve has supersolid phase 
around the 1/2-plateau and the phase transition between the normal BEC and 
the supersolid phase is of second-order.  
All curves in BEC and supersolid phase is convex down 
because of the 3-point interaction $\gamma$ in (\ref{Eeff}).
\label{Fig:magcurve}}
\end{center}
\end{figure}

The critical field $H_{\text{c2}}$ where the saturation occurs is given 
by $(\partial E_{\text{eff}}/\partial \VEV{\sigma^z})_{\VEV{\sigma^z}=-1}=0$ 
after substituting $\VEV{\sigma^{z\prime}}=\VEV{\sigma^z}$, i.e.
\begin{equation}
\begin{split}
2H_{\text{c2}}& =J_1+2J_2-\mu-4t+4J_{\text{eff}1}+4J_{\text{eff}2}\\
& -2L_{\text{eff}}^{(1)} -8L_{\text{eff}}^{(3)}-8L_{\text{eff}}^{(4)}
-2L_{\text{eff}}^{(6)}  \; .
\end{split}
\label{Hc2}
\end{equation}
We minimized $E_{{\rm eff}}$ numerically and we found that the energy 
in the ``stripe'' case was always equal to or smaller than that 
in the ``checkerboard'' case. We show various types of magnetization 
curves obtained in this way in FIG.~\ref{Fig:magcurve}.
In FIG.~\ref{Fig:magsouzu}, we also classified the parameter regions 
(in the $(J_1/J_2,\lambda)$-plane) according to the qualitative behavior
of the magnetization curve.   
There appears (i) the normal BEC phase, (ii) the ``striped'' supersolid phase 
and (iii) the ``striped'' 1/2-plateau.  
At the 1/2-plateau, the pseudo-spins $\sigma$ are ordered in a collinear 
manner $\VEV{\sigma^z}=1$ and $\VEV{\sigma^{z\prime}}=-1$ 
(see FIG.~\ref{Fig:sublattice}).  

The magnetization curve in the BEC and the supersolid phase is convex 
down because of 3-point interaction $\gamma$ in (\ref{Eeff}) 
which breaks the particle-hole symmetry.  
The ``striped'' supersolid phase 
always appears around the 1/2-plateau and the width of 
the supersolid phase appearing on the left of the 1/2 plateau 
is broader than that on the right due to the convex down character.
The equivalent Hamiltonian (\ref{Heff}) without the 3-point interactions  
has been investigated by using the mean field theory\cite{SSmean} and Monte-Carlo simulations\cite{SSmean,Batrouni-Scalettar}. 
They found that the ``striped'' supersolid phase around the 1/2-plateau 
is stable\cite{Batrouni-Scalettar}.  
Therefore, our result that the supersolid phase exists may be correct 
beyond the mean-field approximation, since the 3-point interaction 
in (\ref{Heff}) is weak.
There are other models accompanied by the supersolid phase, e.g. spin dimer XXZ model\cite{dimerSS}, spin-1/2 XXZ model on the triangular lattice\cite{triSS}, etc.

\section{Comparison with the experimental data of ${\bf (CuCl)LaNb_2O_7}$}
\label{sec:CuCl}
In this section, we compare our results with the experimental data 
obtained for $\text{(CuCl)LaNb}_{2}\text{O}_{7}$.  
Since we have three parameters $J_1$,$J_2$ and $\lambda$, three 
experimental inputs in principle determine the set of coupling
constants. 
Then, we use those values of coupling constants to compare 
the magnetization curve of our model with the experimental 
one\cite{Kageyama-05-2}.  

We use the triplet gap $E^{-}_{t}({\bf k}={\bf 0})=26.7\,\text{K}$ 
observed in inelastic neutron scattering\cite{Kageyama-05-1}, 
the lower critical field $H_{\text{c1}}=10.3\text{T}$ 
(or $15.0\text{K}$ if $g=2.17$ is used), 
which marks the onset of magnetization, 
and the saturation field\cite{Kageyama-05-2} 
$H_{\text{c2}}=30.1\text{T}$ $(43.7\text{K})$ as the experimental input.     

\begin{figure}[H]
\begin{center}
\includegraphics[scale=0.77]{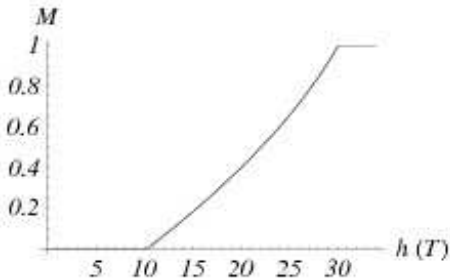}
\caption{Magnetization curve obtained from (\ref{Eeff}) 
by using the parameter set (\ref{CuClpara}).
\label{Fig:CuCl}}
\end{center}
\end{figure}

The triplet gap has been calculated in sec.~\ref{sec3} and are given 
by eq.(\ref{AP2:tmin}) or (\ref{AP2:tmind}).  
In sec.~\ref{sec:magnet}, we have obtained the critical field 
$H_{\text{c1}}$ (eq.(\ref{tatiagari})) and $H_{\text{c2}}$ (eq.(\ref{Hc2})).  
We compare these results with the experimental ones to determine 
two exchange couplings $J_{1}$, $J_{2}$ and the distortion 
parameter $\lambda$.   
The result is:
\begin{equation}
J_1=-140{\rm K} \ ,\ \ J_2=87{\rm K}\ ,\ \ \lambda=0.46\ ,\label{CuClpara}
\end{equation}
where we have used (\ref{AP2:tmind}) for the excitation energy of the
triplet. 
The magnetization curve for the ratio $J_1/J_2=-1.6$ and 
the distortion $\lambda=0.46$ obtained above 
is shown in FIG.~\ref{Fig:CuCl} (see FIG.~\ref{Fig:magsouzu}).
This curve is similar to that obtained in the high-field magnetization 
measurement\cite{Kageyama-05-2} except for the little convex down 
character.   

However, a remark is in order here.  
Recent NMR experiments\cite{Yoshida-07} suggest    
the displacement patterns of $\text{Cl}^{-}$ which yield 
different magnetic interactions from what have been assumed here.  
In particular, the system does not have any explicitly tetramerized 
structure (see FIG.\ref{fig:plafull}), 
although $\text{(CuCl)LaNb}_{2}\text{O}_{7}$ has period 2 
both in the $a$- and the $b$ direction.    
Therefore, our results should not be taken literally.  
Instead, our plaquette model should be thought of as one of 
the simplest Hamiltonians realizing the BEC of magnon bound 
states which is applicable to much wider class of systems 
including our simple $J_{1}$-$J_{2}$ model.  
\section{Conclusion}
\label{sec:Conclusion}
Motivated by the recent discovery of a new two-dimensional 
spin-gap compound $\text{(CuCl)LaNb}_{2}\text{O}_{7}$, 
we have studied spin-1/2 $J_1$-$J_2$ model with a plaquette structure. 
For the small inter-plaquette interactions, i.e., for small $\lambda$, 
there exists a finite spin gap over the spin-singlet ground state.  

We have computed the excitation energy 
of the triplets and the quintet in section \ref{sec3} in two 
different methods.  If the gap of the lowest excitation closes, 
the corresponding particle condenses and a phase transition occurs 
from a paramagnetic phase to magnetically ordered phases.  
For the case of ferromagnetic $J_{1}$ considered here, 
we have two possibilities.  For relatively small $|J_{1}|/J_{2}$, 
the triplet particles (${\bf p}$ and ${\bf q}$) condense and 
generically we may expect CAF appears after the condensation 
(see FIG.~\ref{fig:souzuG}).   

For larger values of $|J_{1}|/J_{2}$, however, 
the quintet excitation matters and we may have various phases.  
In the situation of relevance, we have either a usual ferromagnetic 
phase or a less conventional spin-nematic phase.  
One of these phases is selected by magnetic interaction among 
the quintet particles.  We have derived an effective Hamiltonian 
governing the magnetic part by using the second-order perturbation 
and mapped out the magnetic phase diagram (see FIG.~\ref{fig:nf}).  
A mean-field calculation predicted a finite window of the spin-nematic 
phase (green region in FIG.~\ref{fig:nf}) in agreement with 
recent numerical results\cite{Shannon-06} obtained for $\lambda=1$. 
From the properties of the condensing particle, we found the nematic
order for $-2.33\leq J_1/J_2\leq -1.91$ in the homogeneous ($\lambda=1$) 
$J_1$-$J_2$ model in section \ref{sec:phase}.  
We remark that our effective Hamiltonian is closely related to that 
for $F=2$ cold atoms in optical lattices
\cite{Barnett,Zhou-Semenoff,Lewenstein-review}.   

We have studied the magnetization process in section \ref{sec:magnet}.  
In the region of interest, magnetization is carried by spin-2 particles, 
which should be identified with a tightly-bound magnon pair 
({\em magnon molecule}),and we have constructed 
an effective hardcore boson (or, pseudo spin-1/2) model 
for these spin-2 particles.   
By using a mean-field ansatz, we have determined the ground state 
of the above effective Hamiltonian as a function of the external field 
$h$.  
We have found three different phases: (i) the normal BEC phase, 
(ii) the `striped' supersolid phase and 
(iii) the `striped' 1/2-plateau. 
In the normal BEC phase, the transverse magnetization $\VEV{S^{\pm}}$ 
vanishes unlike the conventional BEC in the spin-dimer model\cite{Tanaka-01}.  

We have compared the results obtained for our $J_{1}$-$J_{2}$ model 
with the experimental data of $\text{(CuCl)LaNb}_{2}\text{O}_{7}$ 
in section \ref{sec:CuCl}.   
Although we have found that our model could qualitatively explain 
the magnon gap in the inelastic neutron scattering 
experiments\cite{Kageyama-05-1} and the magnetization 
curve\cite{Kageyama-05-2}, the structure suggested 
by NMR measurements\cite{Yoshida-07}  
is inconsistent with our tetramerized $J_{1}$-$J_{2}$ model and 
this agreement should not be taken literally.   
Nevertheless, we hope that our scenario `molecular spin-BEC' 
based on a simple $J_{1}$-$J_{2}$ model will capture the basic physics 
which underlies the magnetism of the compound 
$\text{(CuCl)LaNb}_{2}\text{O}_{7}$.  
\section*{ACKNOWLEDGMENTS}
We would like to thank H.~Kageyama for sharing his unpublished results, many helpful discussions, and comments on the manuscript. We are also grateful to A.~Kitada and T.~Miki for many useful discussions, and Mike Zhitomirsky for careful reading of the manuscript. This work is supported by the
Grant-in-Aid for the 21st Century COE of Education, Culture,
Sports, Science and Technology (MEXT) of Japan.

\appendix
\section{Diagonalization of Hamiltonian by Bogoliubov transformation}
\label{sec:appendix-1}
For convenience, we briefly summarize the method of Bogoliubov transformation. We want to diagonalize
\begin{equation}
H=\sum_{{\bf k}}
{\bf v}^{\dagger}_{\bf k}A({\bf k}){\bf v}_{\bf k}\ ,\label{AP1:1}
\end{equation}
where
\begin{subequations}
\begin{align}
{\bf v}_{{\bf k}} &=
\left(
\begin{array}{c}
p_{{\bf k}}\ ,\ q_{{\bf k}}\ ,\ p^{\dagger}_{-{\bf k}}\ ,\ q^{\dagger}_{-{\bf k}}\\
\end{array}
\right) ^{{\rm T}}\ ,\\
A({\bf k})&=\left(
\begin{array}{cccc}
a ({\bf k}) & b ({\bf k}) & c ({\bf k}) & d ({\bf k})\\
b ({\bf k}) & e ({\bf k}) & b ({\bf k}) & d ({\bf k})\\
c ({\bf k}) & b ({\bf k}) & a ({\bf k}) & b ({\bf k})\\
b ({\bf k}) & d ({\bf k}) & b ({\bf k}) & e ({\bf k})\\
\end{array}
\right) \ .\label{AP1:A}
\end{align}
\end{subequations}
Now $p,q$ are boson operators, and $A({\bf k})=A(-{\bf k})$. We introduce Bogoliubov transformation
\begin{equation}
L_{{\bf k}}{\bf v}_{{\bf k}}={\bf v}^{\prime}_{{\bf k}},\ \ \ 
{\bf v}^{\prime}_{{\bf k}}=
\left(
\begin{array}{c}
p^{\prime}_{{\bf k}}\ ,\ q^{\prime}_{{\bf k}}\ ,\ p^{\prime\dagger}_{-{\bf k}}\ ,\ q^{\prime\dagger}_{-{\bf k}}\\
\end{array}
\right) ^{{\rm T}}\ ,\label{AP1:L}
\end{equation}
where $L$ is $4\times 4$ real matrix, and
\begin{equation}
\begin{split}
L_{11}^* ({\bf k}) &=L_{33}(-{\bf k}),\ \ L_{12}^* ({\bf k})=L_{34}(-{\bf k})\ ,\\
L_{13}^* ({\bf k}) &=L_{31}(-{\bf k}),\ \ L_{14}^* ({\bf k})=L_{32}(-{\bf k})\ ,\\
& \vdots \label{AP1:cd1}
\end{split}
\end{equation}
\begin{equation}
\begin{split}
[ p^{\prime},p^{\prime\dagger} ]& =[L_{1\mu}v_\mu,L_{1\nu} v_\nu^{\dagger}]
=L_{11}^2 +L_{12}^2 -L_{13}^2 -L_{14}^2 =1\ ,\\
[ p^{\prime},q^{\prime\dagger} ]& =L_{11}L_{21} +L_{12}L_{22} -L_{13}L_{23} -L_{14}L_{24} 
=0\ ,\\
[ p^{\prime\dagger},p^{\prime} ]& =L_{31}^2 +L_{32}^2 -L_{33}^2 -L_{34}^2 
=-1\ ,\\
& \vdots
\end{split}
\label{AP1:cd2}
\end{equation}
where the summation over repeated indices is implied. By using
\begin{eqnarray}
g=
\left(
\begin{array}{cccc}
1 & 0 & 0 & 0 \\
0 & 1 & 0 & 0 \\
0 & 0 & -1 & 0 \\
0 & 0 & 0 & -1 \\
\end{array}
\right),
\end{eqnarray}
and
\begin{eqnarray}
\epsilon_i=
\left\{
\begin{array}{l}
\epsilon=1\ {\rm for}\ i=1,2\\
\epsilon=-1\ {\rm for}\ i=3,4\\
\end{array}
\right. \ ,
\end{eqnarray}
the condition (\ref{AP1:cd2}) can be rewritten as
\begin{eqnarray}
{\bf l}_i\cdot {\bf l}_j=g_{\mu\nu}l_i^\mu l_j^\nu=\epsilon_i \delta_{ij},\ \ 
\label{beq}
L=
\left(
\begin{array}{c}
{\bf l}_1^{{\rm T}}\\
{\bf l}_2^{{\rm T}}\\
{\bf l}_3^{{\rm T}}\\
{\bf l}_4^{{\rm T}}\\
\end{array}
\right)\ .
\end{eqnarray}
where ${\bf l}$ is a 4-dimensional column vector and the summation over $i$ is not taken in (\ref{beq}). $g_{\mu\nu}$ can be considered as a metric. 
The condition (\ref{AP1:cd1}) can be rewritten as
\begin{eqnarray}
{\bf l}_{(3,4)}(-{\bf k})=
\left(
\begin{array}{c}
{\bf r}^{d*}_{(1,2)}({\bf k})\\
{\bf r}^{u*}_{(1,2)}({\bf k})\\
\end{array}
\right)\ ,\label{Lkkai}
\end{eqnarray}
where ${\bf l}_i=
\left(
\begin{array}{c}
{\bf r}_i ^{u{\rm T}}\ ,\ {\bf r}_i ^{d{\rm T}}\\
\end{array}
\right) ^{{\rm T}}$
and ${\bf r}$ is a 2-dimensional column vector. We denote $A$ in (\ref{AP1:A}) as $A_{\mu\nu}$ and $\tilde{A}\equiv gA$ as $A^{\mu}_{\ \nu}=g^{\mu\rho}A_{\rho\nu}$. Then, with regard to eigenvectors satisfying $A^{\mu}_{\ \nu}u^\nu_a=a u^\mu_a\ ,\ A^{\mu}_{\ \nu}u^\nu_b=b u^\mu_b$, we obtain
\begin{equation}
\begin{split}
g_{\mu\nu}u_a^\mu u_b^\nu=u_{a\mu}u^{\mu}_b&=\frac{1}{a}u_{a\mu}A^{\mu}_{\ \nu}u_b^\nu\ ,\\
  &=\frac{1}{b}u_{a\mu}A^{\mu}_{\ \nu}u_b^\nu\ ,
\end{split}
\end{equation}
since $A_{\mu\nu}$ is symmetric. Therefore, if $a\neq b$, $u_{a\mu}u^{\mu}_b=0$, i.e. eigenvectors of a different eigenvalue are orthogonal each other. We define
\begin{equation}
L^{\prime} =
\left(
\begin{array}{cccc}
{\bf l}^\prime_1 & {\bf l}^\prime_2 & {\bf l}^\prime_3 & {\bf l}^\prime_4\\
\end{array}
\right)^{\rm T}
\end{equation}
where {\bf l}$^{\prime}_{i}(i=1\sim4)$ are eigenvectors of $A^\mu_{\ \nu}$, and 
\begin{eqnarray}
{\bf l}^{\prime2}_{(1,2)}=1,\ \ \ 
{\bf l}^{\prime2}_{(3,4)}=-1\ .
\end{eqnarray} 
Then, $L^\prime g L^{\prime {\rm T}}=g$. Therefore,
\begin{equation}
L^{\prime {\rm T}}gL^{\prime}g=I\ ,\label{AP1:koutou}
\end{equation}
where $I$ is identity matrix. Now we can write
\begin{equation}
A=\left(
\begin{array}{cccc}
A_1 & A_2\\
A_2 & A_1\\
\end{array}
\right)\ ,
\end{equation}
where $A_{1,2}$ are $2\times 2$ matrix, and $A({\bf k})=A(-{\bf k})$. Therefore, we can take {\bf l}$^{\prime}_{i}(i=1\sim4)$ satisfying (\ref{Lkkai}). Defining $\Omega_{i}$ as the eigenvalue of {\bf l}$^{\prime}_{i}(i=1\sim4)$, this leads to
\begin{eqnarray}
\Omega_1=-\Omega_3,\ \ \ \Omega_2=-\Omega_4\label{omegar}
\end{eqnarray}
From (\ref{AP1:koutou}), (\ref{AP1:1}) reduces to
\begin{eqnarray}
{\bf v}^\dagger A {\bf v} &=& {\bf v}^\dagger g^2 A{\bf v}\nonumber\\
&=& {\bf v}^{\prime\dagger} L^{\prime}g\tilde{A}
L^{\prime{\rm T}} {\bf v} ^\prime\nonumber\\
&=& {\bf v}^{\prime\dagger}L^{\prime}g
\left(
\begin{array}{cccc}
\Omega_1{\bf l}^{\prime}_1 & \Omega_2{\bf l}^{\prime}_2 & \Omega_3{\bf l}^{\prime}_3 & \Omega_4{\bf l}^{\prime}_4 \\
\end{array}
\right)
{\bf v}^\prime\nonumber\\
&=& {\bf v}^{\prime\dagger}
\left(
\begin{array}{cccc}
\Omega_1 & 0 & 0 & 0 \\
0 & \Omega_2 & 0 & 0 \\
0 & 0 & -\Omega_3 & 0 \\
0 & 0 & 0 & -\Omega_4 \\
\end{array}
\right)
{\bf v}^\prime \label{tkm}
\end{eqnarray}
where ${\bf v}^{\prime} =gL^{\prime} g {\bf v},\ \tilde{A}=gA$, and {\boldmath $k$} is omitted. Comparing to (\ref{AP1:L}), we obtain
\begin{eqnarray}
L=gL^{\prime}g\ .
\end{eqnarray}
This $L$ satisfies (\ref{beq}) and (\ref{Lkkai}). The eigenvalues of $\tilde{A}$ (see eq. (\ref{AP1:A})) are given by
\begin{equation}
\begin{split}
& \Omega_{(\pm ,\pm)}= \pm [ a^2-c^2-d^2+e^2 \pm \{ \left(-a^2+c^2+d^2-e^2\right)^2\\
& \ \ +4 (a-d) (c-e) \left(-4 b^2+a c+c d+a e+d   e\right) \}^{\frac{1}{2}} ]^{\frac{1}{2}}\ .
\end{split}
\end{equation}
We note that\\
    $\Omega_{1,2}$ are the eigenvalues of the eigenvectors whose norm is positive,\\
    $\Omega_{3,4}$ are the eigenvalues of the eigenvectors whose norm is negative,\\
and (\ref{omegar}) is satisfied. Moreover, using the boson's commutation relation, $H$ reduces to 
\begin{subequations}
\begin{align}
& H=\sum_{{\bf k}}\{
\omega_1({\bf k})p^{\prime\dagger}_{{\bf k}}p^{\prime}_{{\bf k}}
+\omega_2({\bf k}) q^{\prime\dagger}_{{\bf k}}q^{\prime}_{{\bf k}}\}
+E_G\ ,\\
& E_G=\sum_{{\bf k}}\frac{\left\{\omega_1({\bf k}) +\omega_2({\bf k})\right\}}{2}\ .
\end{align}
\end{subequations}
where $\omega_{1,2}=2\Omega_{1,2}$.

\section{How to parametrize general spin-S states}
\label{classify}
In this section, we briefly summarize the method of parametrizing 
arbitrary spin-2 states used in 
a mean-field calculation of section~\ref{sec:phase}.  
The method is based on a geometrical representation of the spin-$S$ states 
used by Bacry\cite{Bacry} and Barnett {\em et al.}\cite{Barnett}.   
Since our model has rotational symmetry, 
the mean-field energy has a trivial degeneracy with respect to 
the global rotation of the spin states.  
To mod out this degeneracy and find only essentially 
different solutions, this geometric method is quite efficient.  

First we introduce the maximally polarized spin-$S$ state 
({\em spin coherent state}) $\ket{\hat{\boldsymbol{\Omega}}}$ 
which is pointing the direction of
\[
\hat{\boldsymbol{\Omega}}
=(\cos\phi\sin\theta,\sin\phi\sin\theta, \cos\theta) \; ,
\] 
i.e. 
$({\bf S}{\cdot}\hat{\boldsymbol{\Omega}})\ket{\hat{\boldsymbol{\Omega}}}
=S\ket{\hat{\boldsymbol{\Omega}}}$. 
If we introduce the Schwinger boson operators $\hat{a}_{+}$ 
($\hat{a}_{-}$) which destroys a spin parallel (anti-parallel) 
to the $z$-direction, the operator 
which creates a spin parallel to the $\hat{\boldsymbol{\Omega}}$-%
direction is given by
\begin{equation}
\hat{v}^\dagger=u\hat{a}_+^\dagger +v\hat{a}_-^\dagger\ ,
\end{equation}
where
\begin{equation}
u=\be^{-i\frac{\chi}{2}}\be^{-i\frac{\phi}{2}}\cos \frac{\theta}{2}\ ,\quad
v=\be^{-i\frac{\chi}{2}}\be^{i\frac{\phi}{2}}\sin \frac{\theta}{2}
\end{equation}
and $\chi$ is an arbitrary gauge function.  
By using $\hat{v}^{\dagger}$, the coherent state 
$\ket{\hat{\boldsymbol{\Omega}}}$ can be written simply as
\begin{equation}
\begin{split}
\ket{\hat{\boldsymbol{\Omega}}}& 
=\frac{1}{\sqrt{(2S)!}}(\hat{v}^\dagger)^{2S}\ket{0}\\
& =v^{2S}\sum_{p=0}^{2S}\sqrt{_{2S}\text{C}_p}
\left(\frac{u}{v}\right)^p \ket{S:Sz=p-S}\ ,
\end{split}
\end{equation}
where the combinatorial symbol $_{2S}\text{C}_p$ is defined by 
$_{2S}\text{C}_p\equiv (2S)!/((2S{-}p)!p!)$.  

Next, we introduce a complex number 
$\zeta=(u/v)^\ast=e^{i\phi}\cot \frac{\theta}{2}$ and 
the corresponding unnormalized ket $\ket{\zeta}$:
\begin{equation}
\ket{\zeta}\equiv\sum_{p=0}^{2S}\sqrt{_{2S}\text{C}_p}(\zeta^\ast)^p
\ket{S:p-S}\ .
\end{equation}
We note that the vector $\hat{\boldsymbol{\Omega}}$ rotates 
on the unit sphere $\text{S}^{2}$, when the SU(2) rotation operator 
$\hat{D}$ acts on $\ket{\zeta}$.  
We denote an arbitrary spin-$S$ state by 
$\ket{A}=\sum_{p=0}^{2S}A_p\ket{S:p-S}$. 
Then, it is convenient to introduce the following `wave function' 
which is in a one-to-one (except for an unphysical overall phase factor)
correspondence with $\ket{A}$ under 
the condition $\sum |A|^{2}=1$:
\begin{equation}
\begin{split}
P_s(\zeta) & \equiv \langle \zeta |A \rangle \\
& =\sum_{p=0}^{2S}\sqrt{_{2S}\text{C}_p}\,A_p \,\zeta^p \\
& = A_{2S}\prod_{i=1}^{2S}(\zeta-\alpha_i)\ ,\\
\end{split}
\label{Pzeta}
\end{equation}
where $\alpha_i$s are the $2S$ roots of $P_s(\zeta)=0$ and 
are parametrized as $\alpha_{i}=e^{i\phi_i}\cot \frac{\theta_i}{2}$.  
If the degree $\text{deg}$ of the above polynomial is smaller than $2S$, 
$(2S-\text{deg})$ roots of $P_{S}(\zeta)$ are at the infinity 
($\theta_{i}=0$ or the north pole).  
Since the stereographic projection uniquely maps 
a set of $2S$ complex roots $\{\alpha_{i}\}$ onto a set of 
$2S$ points on a two-dimensional sphere $\text{S}^{2}$,  
we can parametrize arbitrary spin-$S$ states by specifying 
$2S$ points on a sphere.  
 
If $A_{2S}=0$, the limit $A_{2S}\rightarrow 0,\ \alpha_{j}=O(1/A_{2S})$ 
for any $j$ must be taken ($\theta_j\rightarrow 0$).  
In the case of spin-2, $A^\prime_i=A_i/A_{2S}$ is given 
in terms of four complex numbers $\{\alpha_{i}\}$ by
\begin{equation}
\begin{split}
A^\prime_{0}& =\alpha_1\alpha_2\alpha_3\alpha_4\ ,\\
A^\prime_{1}& =-\frac{\alpha_1\alpha_2\alpha_3 
+\alpha_1\alpha_2\alpha_4+\alpha_1\alpha_3\alpha_4
+\alpha_2\alpha_3\alpha_4}{2}\ ,\\
A^\prime_{2}& =\frac{\alpha_1\alpha_2+\alpha_1\alpha_3
+\alpha_1\alpha_4+\alpha_2\alpha_3+\alpha_2\alpha_4
+\alpha_3\alpha_4}{\sqrt{6}}\ ,\\
A^\prime_{3}& =-\frac{\alpha_1+\alpha_2+\alpha_3+\alpha_4}{2}\ ,\\
A^\prime_{4}& =1\ ,
\end{split}
\label{eqn:AbyAlpha-1}
\end{equation}
and hence the coefficients $\{A_{i}\}$ read
\begin{equation}
A_{i}=\frac{e^{i\phi}}{\sqrt{\sum_{i^\prime=0}^{4}
|A^\prime_{i^\prime}|^2}}A^\prime_{i} \quad (i=0,\ldots,4)\ ,
\label{eqn:AbyAlpha-2}
\end{equation}
where $\phi$ is the phase of $A_{4}$.  
Therefore, as has been described above, 
arbitrary spin-2 states are parametrized by 
a set of four unit vectors and an overall phase factor.  
The rotational symmetry enables us to further reduce the number 
of free parameters by fixing $\alpha_{1}$ and 
$\alpha_{2}$ as:
\begin{equation}
\alpha_1=1, \ \alpha_2=e^{i\phi_2},\ \alpha_3 
=e^{i\phi_3}\cot \frac{\theta_3}{2},\ \alpha_4 
=e^{i\phi_4}\cot \frac{\theta_4}{2} \; .
\label{eqn:AbyAlpha-3}
\end{equation}
Equations (\ref{eqn:AbyAlpha-1})-(\ref{eqn:AbyAlpha-3}) express 
arbitrary (except for global rotation) spin-2 states in terms of 
five free parameters.    
\section{Relation among the expectation values of spin-S operators}
\label{sec:AP3}
There exists a simple relation among the expectation values 
of spin-$S$ operators.  By spin-$S$ operators, here we mean  
all independent (traceless) polynomials made up of the usual spin-$S$ 
operators ${\bf S}$.   
The spin 1 case has been considered by Chen and Levy\cite{Chen-Levy} 
in the context of spin-nematic order.  An arbitrary spin-$S$ ket 
is written as
\begin{equation}
{\bf z}=\left(
\begin{array}{c}
z_1\\
z_2\\
\vdots\\
z_{2S+1}
\end{array}
\right)\ ,
\end{equation}
where $\sum_i^{2S+1} |z_i|^2=1$.  It is convenient to consider 
the Lie group $\text{SU}(2S+1)$ which naturally acts on the above 
$(2S{+}1)$-dimensional space. 
Let us denote the generators $T^a$ $(a=1,\ldots,(2S+1)^2{-}1)$ of $SU(2S+1)$ 
and normalize them as
\begin{equation}
{\rm Tr}(T^aT^b)=\frac{1}{2}\delta^{ab}\ .
\end{equation}
Then, they satisfy 
\begin{equation}
\begin{split}
& \sum_a T^a_{ij}\, T^a_{kl}=\frac{1}{2}\left(
\delta_{il}\delta_{jk}-\frac{1}{2S+1}\delta_{ij}\delta_{kl}
\right) \\
& (i,j,k,l=1,\ldots,2S+1) \ .
\end{split}
\end{equation}
Using this relation, we obtain
\begin{equation}
\begin{split}
\sum_a \VEV{T^a}^2 & =(z^\dagger_iT_{ij}^a z_j)(z^\dagger_k T^a_{kl}z_l)\\
& =\frac{1}{2}\left\{ z_i^\dagger z_j z_j^\dagger z_i
-\frac{1}{2S+1}(z_i^\dagger z_i)(z_k^\dagger z_k)\right\}\\
& =\frac{S}{2S+1}\ ,
\end{split}
\end{equation}
where the summation over repeated indices is implied. In spin-1/2 
($SU(2)$) case, $T^a$ can be written as $\frac{1}{2}\sigma^a$, 
where $\sigma$ is Pauli matrix. Therefore, this relation can be written as
\begin{equation}
\sum_{i=x,y,z} \VEV{\sigma_i}^2=1\ ,\label{sigmaI}
\end{equation}
\begin{widetext}
\section{Expressions of omitted equations}
\label{sec:AP2}
\underline{Section \ref{sec:2per}}:\\
The elements of the second-order hopping matrix (eq. (\ref{acb})) are given by:
\begin{equation}
\begin{split}
a({\bf k}) \equiv 
-\lambda \frac{J_{2}}{2}f_{+}({\bf k})
-\lambda^{2}\frac{J_{2}}{4}f_{+}({\bf k}) +\lambda^{2}\left\{\frac{-4J_{1}^{5}+3J_{1}^{4}J_{2}
+ 24J_{1}^{3}J_{2}^{2}-25J_{1}^{2}J_{2}^{3}+28J_{1}J_{2}^{4}
- 28J_{2}^{5}}{8J_{2}(J_{1}-2J_{2})(J_{1}-J_{2})(J_{1}+2J_{2})}
-\frac{J_{2}}{4}\right\}\ ,
\end{split}
\label{AP2:p}
\end{equation}
\begin{equation}
\begin{split}
b({\bf k})\equiv
-\lambda \frac{J_{2}}{2}f_{-}({\bf k})
-\lambda^{2}\frac{J_{2}}{4}f_{-}({\bf k})  +\lambda^{2}\left\{
\frac{-4J_{1}^{5}+3J_{1}^{4}J_{2}+24J_{1}^{3}J_{2}^{2}
-25J_{1}^{2}J_{2}^{3}+28J_{1}J_{2}^{4}
-28J_{2}^{5}}{8J_{2}(J_{1}-2J_{2})(J_{1}-J_{2})(J_{1}+2J_{2})}
-\frac{J_{2}}{4}\right\}\ .
\end{split}
\label{AP2:q}
\end{equation}
\begin{eqnarray}
f_{\pm}({\bf k})\equiv\cos k_{x}
+\cos k_{y}+\cos(k_{x}\pm k_{y})\ .
\end{eqnarray}
The second-order energy shift for the singlet ground state is calculated 
as(see eq. (\ref{Etpm})):
\begin{equation}
\Delta E_s\equiv-\frac{3 \lambda ^2 \left(2 J_1^2+3 J_2^2\right)}{8
   J_2}\ .
\label{AP2:Es}
\end{equation}
The excitation gap of triplets from the second-order perturbation 
is given by:
\begin{eqnarray}
\Delta_{t}\equiv E_t^-({\bf k}={\bf 0}) =J_{2} - \lambda\frac{3}{2}J_2
+\lambda^2\frac{2 J_1^5-3 J_2 J_1^4+J_2^2 J_1^3-2 J_2^3 J_1^2+24 J_2^4J_1
-24 J_2^5}{8 J_2 \left(J_2-J_1\right) \left(2 J_2-J_1\right) 
\left(J_1+2 J_2\right)}\ .\label{AP2:tmin}
\end{eqnarray}
The modified excitation gap of triplets which is free from the pole 
$J_1=2J_2$ is given by:
\begin{eqnarray}
\Delta_{t,\text{mod}}\equiv 
E_{t,{\rm mod}}^{-}({\bf k}={\bf 0})= J_2-\lambda\frac{3}{2}J_2+\lambda^2
\frac{25  J_1^2-65 \ J_2^2}{144 J_2}\ .\label{AP2:tmind}
\end{eqnarray}
The excitation energy of quintet from the second-order perturbation 
is given by:
\begin{eqnarray}
E_q({\bf k})=
J_1+2 J_2  +\lambda ^2\left\{
\frac{-14 J_1^3+4 J_2 J_1^2+15 J_2^2 J_1-6 J_2^3}{8 J_1 \left(2 J_1-J_2\right)}
+\frac{   \left(J_1^2+J_2^2\right)}{4 J_1}(\cos k_x+\cos k_y)\right\}
\ .\label{AP2:quin}
\end{eqnarray}
\underline{Section \ref{sec:phase}}:\\
The parameters of the effective Hamiltonian (\ref{Hqumean}) 
where the quintet condenses are given by:
\begin{equation}
J_{{\rm qu}1}=\frac{1}{8} \lambda  \left(J_1+J_2\right)
-\frac{\lambda ^2 \left(59 J_1^4+78 J_2 J_1^3+60 J_2^2
   J_1^2+26 J_2^3 J_1+J_2^4\right)}{576 J_1 \left(3
   J_1^2+4 J_2 J_1+J_2^2\right)}\ .
\end{equation}
\begin{equation}
J_{{\rm qu}2}=\frac{\lambda  J_2}{16}
-\frac{\lambda ^2 J_2^2 \left(133 J_1^2+76 J_2 J_1+7
   J_2^2\right)}{2304 J_1 \left(3 J_1^2+4 J_2
   J_1+J_2^2\right)}\ .
\end{equation}
\begin{equation}
K_{{\rm qu}1}=-\lambda^2 \frac{49 J_1^4+174 J_2 J_1^3
+120 J_2^2 J_1^2+10 J_2^3
   J_1-J_2^4}{2304 J_1 \left(3 J_1^2+4 J_2
   J_1+J_2^2\right)}\ ,\ \ 
K_{{\rm qu}2}=-\lambda^2 \frac{11 J_2^4+68 J_1 J_2^3
+137 J_1^2 J_2^2}{9216 J_1
   \left(3 J_1^2+4 J_2 J_1+J_2^2\right)}\ .
\end{equation}
\begin{equation}
\begin{split}
& L_{{\rm qu}1}^{(1)}=\frac{ \lambda ^2 \left(J_1+J_2\right)}{576},\ 
L_{{\rm qu}1}^{(2)}=0,\ 
L_{{\rm qu}1}^{(3)}=\frac{\lambda ^2 J_2}{1152}\ ,\\
& L_{{\rm qu}1}^{(4)}=-\frac{\lambda ^2 J_2}{1152},\ 
L_{{\rm qu}1}^{(5)}=\frac{\lambda ^2 J_2^2}{4608 J_1},\ 
L_{{\rm qu}1}^{(6)}=-\frac{\lambda ^2 J_2^2 \left(J_2-3 J_1\right)}{4608
   J_1 \left(J_1+J_2\right)},
\end{split}
\end{equation}
\begin{equation}
L_{{\rm qu}2}^{(i)}=-2L_{{\rm qu}1}^{(i)} \ , \quad \text{for } 
i=1,\ldots,6\ .
\end{equation}
\underline{Section \ref{sec:magnet}}:\\
The parameters of the effective Hamiltonian $H_{\text{eff}}$ (\ref{Heff}) 
in the magnetization process are given in a series in $\lambda$ by:
\begin{subequations}
\begin{align}
& -\mu\equiv \frac{\lambda ^2 \left(-14 J_1^3+4 J_2 J_1^2+15 J_2^2
   J_1-6 J_2^3\right)}{8 J_1 \left(2 J_1-J_2\right)}\ .\label{AP2:mu}
\\
& J_{{\rm eff}1}\equiv \lambda\frac{J_1+J_2}{2} 
+\lambda ^2 \frac{\left(6 J_1^4+11 J_2 J_1^3+2 J_2^2
   J_1^2-13 J_2^3 J_1+4 J_2^4\right)}{16 J_1 \left(2
   J_1-J_2\right) J_2}\ .\label{AP2:Je1} 
\\
& J_{{\rm eff}2}\equiv \lambda\frac{J_2}{4}+\lambda^2 
\frac{J_2 \left(6 J_1^2-13 J_2 J_1+4
   J_2^2\right)}{32 J_1 \left(2 J_1-J_2\right)}\ .\label{AP2:Je2}
\end{align}
On top of them, we have three-body (or, three-plaquette) interactions:
\begin{equation}
L_{{\rm eff}}^{(1)}=\lambda^2\frac{\left(J_1+J_2\right)^2}{4 J_2},\ 
L_{{\rm eff}}^{(2)}=0,\ 
L_{{\rm eff}}^{(3)}=\frac{\lambda^2}{8} \left(J_1+J_2\right),\ 
L_{{\rm eff}}^{(4)}=-\frac{\lambda^2}{8} \left(J_1+J_2\right),\ 
L_{{\rm eff}}^{(5)}=0,\ 
L_{{\rm eff}}^{(6)}=\lambda^2\frac{J_2}{8}\ .
\end{equation}
\end{subequations}

The parameters necessary for the mean-field energy (\ref{Eeff}) 
in the external magnetic field depend on the sublattice structures 
assumed in the calculation and are given as follows.  
\begin{enumerate}
\item In the case of ``checkerboard'' sublattice:
\begin{equation}
\begin{split}
\xi& =J_{{\rm eff}1}+J_{{\rm eff}2}+\frac{1}{4}L_{{\rm eff}}^{(1)}
+L_{{\rm eff}}^{(3)}+L_{{\rm eff}}^{(4)}+\frac{1}{4}L_{{\rm eff}}^{(6)}\ ,\\
\alpha_1& =\frac{1}{2}\left(J_{{\rm eff}1}-L_{{\rm eff}}^{(1)}\right), \ \ 
\alpha_2=\frac{1}{4}\left(2J_{{\rm eff}2}+L_{{\rm eff}}^{(1)}
-4L_{{\rm eff}}^{(3)}-4L_{{\rm eff}}^{(4)}-L_{{\rm eff}}^{(6)}\right)\ ,\\
\beta_1& =t,\ \ 
\beta_2=0,\ \ 
\gamma_1=\frac{1}{4}L_{{\rm eff}}^{(1)}+L_{{\rm eff}}^{(3)}
+L_{{\rm eff}}^{(4)} ,\ \ 
\gamma_2=\frac{1}{4}L_{{\rm eff}}^{(6)} \ .
\end{split}
\end{equation}
\item In the case of ``striped'' sublattice:
\begin{equation}
\begin{split}
\xi& =J_{{\rm eff}1}+J_{{\rm eff}2}+\frac{1}{4}L_{{\rm eff}}^{(1)}
+L_{{\rm eff}}^{(3)}+L_{{\rm eff}}^{(4)}+\frac{1}{4}L_{{\rm eff}}^{(6)}\ ,\\
\alpha_1& =\frac{1}{4}\left( J_{{\rm eff}1}+2J_{{\rm eff}2}
-L_{{\rm eff}}^{(1)}-4L_{{\rm eff}}^{(3)}-4L_{{\rm eff}}^{(4)}
-2L_{{\rm eff}}^{(6)}\right),\ \ 
\alpha_2=\frac{1}{4}\left(J_{{\rm eff}1}+L_{{\rm eff}}^{(6)}\right)\ ,\\
\beta_1& =\frac{t}{2},\ \ \beta_2=\frac{t}{2},\ \ 
\gamma_1=\frac{1}{8}\left(L_{{\rm eff}}^{(1)}+8L_{{\rm eff}}^{(3)}
+8L_{{\rm eff}}^{(4)}+2L_{{\rm eff}}^{(6)}\right),\ \ 
\gamma_2=\frac{1}{8}L_{{\rm eff}}^{(1)} \ .
\end{split}
\end{equation}
\end{enumerate}
\end{widetext}


\end{document}